\documentclass[
 reprint,
 showpacs, preprintnumbers,
 amsmath, amssymb,
 aps,
 prl,
 floatfix,
 superscriptaddress
]{revtex4-2}

\usepackage{amsmath}
\usepackage{graphicx}
\usepackage{dcolumn}
\usepackage{bm}
\usepackage{bbm}

\usepackage[utf8]{inputenc}
\usepackage[T1]{fontenc}
\usepackage{mathptmx}
\usepackage{textgreek}
\usepackage{upgreek}
\usepackage{float}
\usepackage{hyperref}

\usepackage{color}

\usepackage{natbib}

\newcommand{\beginsupplement}{%
        \setcounter{table}{0}
        \renewcommand{\thetable}{S\arabic{table}}%
        \setcounter{figure}{0}
        \renewcommand{\thefigure}{S\arabic{figure}}%
				\renewcommand{\theequation}{S.\arabic{equation}}
     }
		
\begin{document}

	\preprint{}

\title{Anisotropic vortex squeezing in synthetic Rashba superconductors: a manifestation of Lifshitz invariants} 



\author{L.~Fuchs}
\affiliation{Institut f\"ur Experimentelle und Angewandte Physik, University of Regensburg, 93040 Regensburg, Germany}
\author{D.~Kochan}
\affiliation{Institut f\"ur Theoretische Physik, University of Regensburg, 93040 Regensburg, Germany}
\affiliation{Institute of Physics, Slovak Academy of Sciences, 84511 Bratislava, Slovakia}
\author{J.~Schmidt}
\affiliation{Institut f\"ur Experimentelle und Angewandte Physik, University of Regensburg, 93040 Regensburg, Germany}
\author{N.~H\"{u}ttner}
\affiliation{Institut f\"ur Experimentelle und Angewandte Physik, University of Regensburg, 93040 Regensburg, Germany}

\author{C.~Baumgartner}
\affiliation{Institut f\"ur Experimentelle und Angewandte Physik, University of Regensburg, 93040 Regensburg, Germany}
\author{S.~Reinhardt}
\affiliation{Institut f\"ur Experimentelle und Angewandte Physik, University of Regensburg, 93040 Regensburg, Germany}

\author{S.~Gronin}
\author{G.~C.~Gardner}
\affiliation{Birck Nanotechnology Center, Purdue University, West Lafayette, Indiana 47907 USA}
\author{T.~Lindemann}
\affiliation{Department of Physics and Astronomy, Purdue University, West Lafayette, Indiana 47907 USA}
\affiliation{Birck Nanotechnology Center, Purdue University, West Lafayette, Indiana 47907 USA}

\author{M.~J.~Manfra}
\affiliation{Birck Nanotechnology Center, Purdue University, West Lafayette, Indiana 47907 USA}
\affiliation{Department of Physics and Astronomy, Purdue University, West Lafayette, Indiana 47907 USA}
\affiliation{School of Materials Engineering, Purdue University, West Lafayette, Indiana 47907 USA}
\affiliation{School of Electrical and Computer Engineering, Purdue University, West Lafayette, Indiana 47907 USA}

\author{C.~Strunk}
\author{N.~Paradiso}\email{nicola.paradiso@physik.uni-regensburg.de}
\affiliation{Institut f\"ur Experimentelle und Angewandte Physik, University of Regensburg, 93040 Regensburg, Germany}
%


\begin{abstract}
Epitaxial superconductor/semiconductor heterostructures combine superconductivity with strong spin-orbit interaction resulting in synthetic Rashba superconductors. The theoretical description of such superconductors involves Lifshitz invariants that are predicted to feature numerous exotic effects with so far sparse experimental evidence. Using a new observable --vortex inductance-- we investigate the pinning properties of epitaxial Al/InAs-based heterostructures. We find a pronounced \textit{decrease} of the vortex inductance with increasing in-plane field which corresponds to a counterintuitive increase of the pinning force. When rotating the in-plane component of the field with respect to the current direction, the pinning interaction turns out to be highly anisotropic. We analytically demonstrate that both the pinning enhancement and its anisotropy are consequences of the presence of Lifshitz invariant terms in the Ginzburg-Landau free energy. Hence, our experiment provides access to a fundamental property of Rashba superconductors and offers an entirely new approach to vortex manipulation.
\end{abstract}

\maketitle

Breaking the inversion symmetry in superconductors has numerous important consequences~\cite{Frigeri2004,BauerSigristBook2012,Smidman2017}. Often it occurs through the Rashba spin-orbit interaction (SOI) which spin-splits the Fermi surface and links the electron spin to the momentum. If the SOI is strong enough to compete with the superconducting pairing, 
it gives rise to a plenty of interesting phenomena, as e.g.~singlet-triplet mixing~\cite{Edelstein1989,GorkovRashba2001}, unconventional pairing~\cite{Fu2008,Fujimoto2008,Zhang2008,Phan2021}, Ising superconductivity~\cite{Xi2016,Lu1353}, magnetochiral resistance~\cite{Wakatsuki2017,Ideue2017,Itahashi2020,Hoshino2018,Tokura2018}, anomalous Josephson effect~\cite{Bezuglyi2002,Buzdin2008,Reynoso2008,Reynoso2012,Yokoyama2014,Shen2014,Konschelle2015,Szombati2016,Assouline2019,Mayer2020b,Strambini2020}, supercurrent diode effect~\cite{Ando2020,Baumgartner2021,baumgartner2021arxiv}, topological superconductivity~\cite{Alicea2010,He2018,Shaffer2020}, and helical 
phases~\cite{Mineev94,Yanase2008,AGTERBERG200313}
with a spatially modulated order parameter.

One possibility to engineer \emph{synthetic} 2D Rashba super\-con\-duc\-tors consists in proximitizing a 2D electron gas (2DEG) with large Rashba SOI by a standard $s$-wave superconductor. This can be realized, e.g., by epitaxially growing an Al film on a shallow InAs quantum well~\cite{Shabani2016,Kjaergaard2016,KjaergaardPRAPPL,MayerAPL2019A}. 
Owing to their non-trivial topological features~\cite{Fu2008,Alicea2010,Potter2011}, such hybrid 2D semiconductor-superconductor heterostructures have been intensely investigated. So far, theory and experiments mainly aimed at exploring the Majorana modes at the edge of topological superconductors, which is enabled by the bulk-boundary correspondence~\cite{HellPRB17,HellPRL17,SuominenPRL2017,NichelePRL2017,PientkaPRX17,Lee2019,Fornieri2019}.
In contrast, experimental signatures of the impact of SOI on the superfluid condensate as such are rather sparse in hybrid systems~\cite{Phan2021}. 


Besides the microscopic description in terms of the Bogoliubov-de Gennes or Gor'kov equations, the effects of SOI and magnetic field on the superconducting condensate can be accounted for phenomenologically
by adding new terms into the underlying Ginzburg-Landau free energy. Such terms---called \emph{Lifshitz invariants}~\cite{Edelstein1996,MineevPRB2008,AGTERBERG200313}---depend on the crystal point-group symmetry~\cite{Smidman2017} and, in the simplest case, they form triple products of magnetic field, linear spatial gradient of the order parameter and a vector specified by SOI. 
The presence of the Lifshitz invariants is theoretically predicted to lead to an anisotropic response of the superflow 
and gives rise to  magnetoelectric effects~\cite{EdelsteinPRL1995,FujimotoPRB2005}, helical phases~\cite{KaurPRL2005,DimitrovaPRB2007,AgterbergPRB2007}, anomalous magnetization~\cite{Levitov1985,Yip2005}, vortex lattice reorientation~\cite{Cameron2019}, and anomalous $\varphi_0$-shift in Josephson junctions~\cite{Buzdin2008}. To our knowledge, there are so far no experimental evidences of the Lifshitz invariant.

\begin{figure*}[tb]
\includegraphics[width=2\columnwidth]{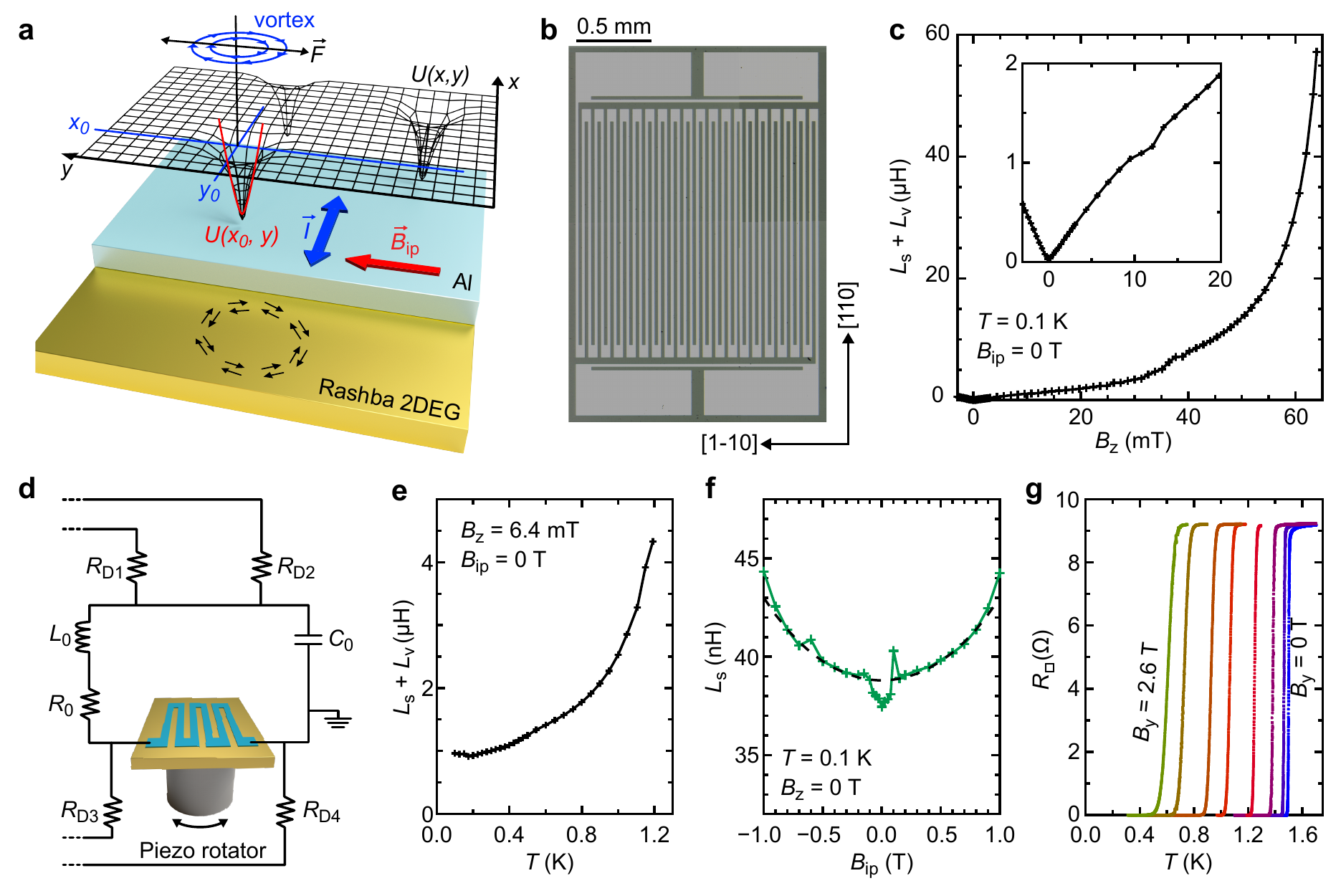}
\caption{\textbf{Vortex inductance as a probe of the pinning potential.} \textbf{a,} Sketch of the device under study. An epitaxial Al film (light blue) proximitizes from the top a shallow InAs quantum well (yellow). The sample is patterned as a 24\,\textmu m-wide and 7.3\,cm-long meander (see micrograph in panel \textbf{b}). The current flows mainly along the $\mathbf{\hat{x}}$ direction, and is subjected to a vortex-generating out-of-plane magnetic field $B_z$ and an in-plane field $\mathbf{B}_{ip}\equiv B_x\mathbf{\hat{x}}+B_y\mathbf{\hat{y}}$ 
at a variable angle $\theta$ with respect to the $\mathbf{\hat{x}}$-axis, which can be controlled. 
The grid represents the vortex free energy $U(x,y)$ for displacement from the pinning centers. An AC current $\mathbf{I}\parallel \mathbf{\hat{x}}$ exerts Lorentz force $\mathbf{F}\parallel \mathbf{\hat{y}}$. The restoring potential in harmonic approximation (small oscillations) is $U(x_0,y)=k_{y}(y-y_0)^2/2$ (red parabola), with $k_y=\partial_y^2U(x_0,y_0)$. 
\textbf{b,} Optical micrograph of the sample. Light grey area corresponds to the Al film, while the dark green ones are etched down to the mesa. 
\textbf{c,} Vortex ($L_v$) plus kinetic ($L_s$) inductance as a function of $B_z$. In our samples $L_s$ ($\approx 40$~nH, see next panel) is negligible compared to $L_v$. The graph shows that by increasing the vortex density, the inductance increases. 
At low fields (inset) the increase is linear Eq.~(\ref{eq:vortind}). At larger vortex densities, the increase is super-linear owing to pair-breaking, which leads to divergence at $B_{c2}=61$~mT. 
\textbf{d,} Measurement scheme: the sample is embedded in a RLC circuit at low temperature and can be rotated with respects to $\mathbf{B}_{ip}$ by means of a piezo rotator. 
\textbf{e,} Measured inductance for $B_z=6.4$\,mT as a function of temperature. 
\textbf{f,} Kinetic inductance vs.~$B_y$ for $B_x=B_z=0$. 
\textbf{g,} Temperature dependence of the sheet-resistance measured at $B_x=B_z=0$ for (right to left) $B_y = 0$, 0.5, 1, 1.5, 2, 2.25, 2.5, 2.6\,T.   
}
\label{fig:firstfig}
\end{figure*}

Vortices can be used to probe the structure of the order parameter  $\Psi(x,y)$, because $|\Psi(x,y)|^2$ near the vortex core is proportional to the potential $U(\mathbf{r})$ that confines a vortex near a point-like pinning site, 
with $\mathbf{r}=(x,y)$ being the vortex displacement from the pinning center at $(x_0,y_0)$, see Fig.~\ref{fig:firstfig}\textbf{a}~\footnote{More precisely, the free energy is a convolution of the defect's pinning potential with the profile of $|\Psi(x,y)|^2$~\cite{BlatterRMP}}. In parabolic approximation the potential $U(\mathbf{r})\,\simeq\, k\,\mathbf{r}^2/2$ 
is characterized only by its curvature $k$~\footnote{Note that in the literature one often finds a related quantity, the so-called Labusch parameter  $k_p=k/d$, where $d$ stands for the film thickness~\cite{Beasley1969,Golosovsky1996}.}.
Driving vortex oscillations around the pinning centers with an AC-current leads to an inductive voltage response, i.e., a vortex inductance
\begin{equation}
    L_{v}\ =\ N_{\square}\,\frac{B_{z}\Phi_0}{k}\;,
    \label{eq:vortind}
\end{equation}
where  
$B_{z}$ is the out-of-plane magnetic field, $\Phi_0=h/(2e)$  the superconducting flux quantum and $N_{\square}=l/w$ being the ratio of length $l$  and width $w$ of the film \cite{Beasley1969,Gittleman1966,Golosovsky1996}.


In this work, we demonstrate 
an unusual, anisotropic \textit{decrease} of the vortex inductance when varying the magnitude and spatial orientation of the in-plane magnetic field.  We interpret this observation as an experimental signature for the so far elusive Lifshitz invariant terms in the Ginzburg-Landau equations for $\Psi(x,y)$. In presence of an in-plane field, an enhancement of the pinning force is observed that reflects an elliptic contraction of the order parameter profile. 
Such pinning enhancement is hard to explain by other known mechanisms, and offers a direct insight into the unusual structure of the order parameter near the vortex cores of Rashba superconductors.



Figure~\ref{fig:firstfig}\textbf{a} shows a schematic of the pinning landscape $U(x,y)$ for a pinned vortex together with the directions of in-plane magnetic field and the AC drive current. A supercurrent in $x$-direction generates a Lorentz force that displaces vortices in $y$-direction from their equilibrium positions. This increases the free energy, producing a restoring force.  For small displacements and low frequencies, pinned vortices thus behave as  underdamped harmonic oscillators~\cite{Beasley1969,Gittleman1966,Golosovsky1996}.

Our synthetic Rashba-superconductor is fabricated starting from a InAs/InGaAs quantum well capped by an epitaxial Al film of nominal thickness $d=7$\,nm~\cite{baumgartner2020}. The Al film induces superconducting correlations in the shallow 2DEG by proximity effect. 
The penetration depth $\lambda$ and the coherence length $\xi$ for the Al/2DEG system at 100\,mK are 227\,nm and 73\,nm, respectively \cite{supplement}.
Using optical lithography and wet etching, we pattern a meander structure, as depicted in Fig.~\ref{fig:firstfig}\textbf{b}. The meander is 24\,\textmu m-wide, which is larger than the Pearl penetration depth $\lambda_{\perp}=2\lambda^2/d=8$\,\textmu m, and a total length of 7.3\,cm, 
resulting in $N_{\square}=3042$ squares. These dimensions are motivated by the need of having at the same time a device in the 2D regime and a large number of squares to increase the measured vortex and kinetic inductance.

The sample holder is mounted on a piezo rotator, whose rotation axis is parallel to the $\mathbf{\hat{z}}$ axis, i.e., perpendicular to the film.  A superconducting coil provides an in-plane magnetic field parallel to the $\mathbf{\hat{y}}$ axis, while an orthogonal pair of coils provides a small out-of-plane field in the $\mathbf{\hat{z}}$ direction. 
The device under study is embedded in a RLC resonant circuit located on the sample holder, see Fig.~\ref{fig:firstfig}\textbf{d}. The circuit, described in Ref.~\cite{baumgartner2020}, allows us to simultaneously measure DC transport characteristics and sample inductance. The latter is deduced from the center frequency shift of the RLC resonance spectrum, which is measured by lock-in detection in the few MHz regime. This is far below the characteristic frequency $\omega_0/2\pi=R_N/2\pi L_v\simeq 4.6$\,GHz ($R_N$ being the normal state resistance)~\cite{Golosovsky1996,Gittleman1966} that separates inductive and dissipative regimes~\footnote{Even a few ohms of AC-resistance (four orders of magnitude less than $R_N$) damp the resonance $Q$ factor below unity.}.
In our inductance measurements we apply a maximum AC excitation of 10~\textmu V to $R_{\text{D1}}$, see Fig.~\ref{fig:firstfig}\textbf{d}. This corresponds to an AC current of 10~nA at low frequency and roughly 350~nA near the resonance, three orders of magnitude less than the critical current. In this regime, all the results here shown are independent from the excitation amplitude.


\begin{figure*}[tb]
\includegraphics[width=\textwidth]{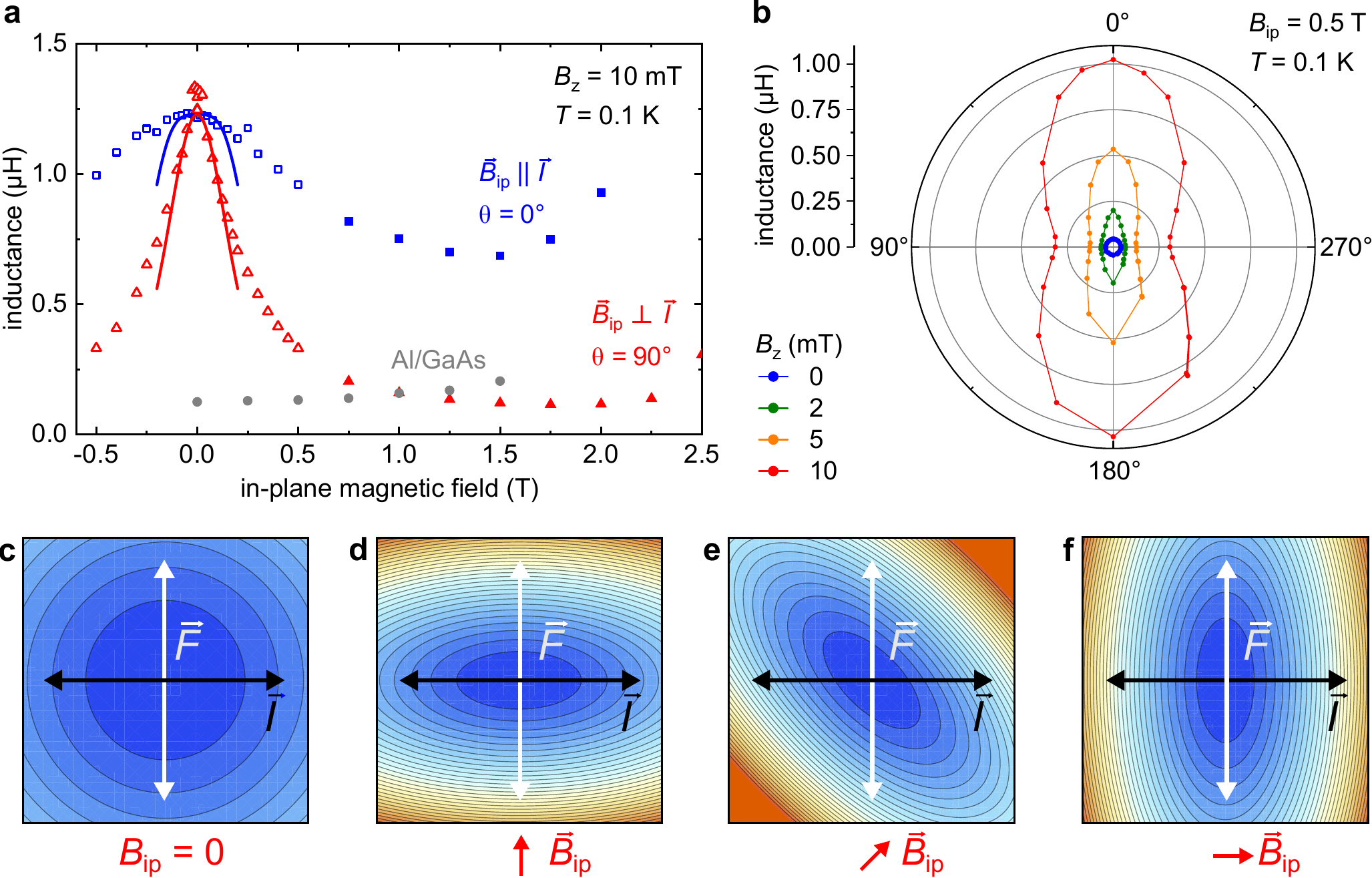}
\caption{ \textbf{Anisotropic vortex squeezing under in-plane field.}
\textbf{a,} Sample inductance measured at $T=0.1$\,K as a function of $B_{ip}$ 
for $\mathbf{B}_{ip}\parallel \mathbf{\hat{x}}$ ($\theta=0^{\circ}$, blue symbols)
and for $\mathbf{B}_{ip}\parallel \mathbf{\hat{y}}$ ($\theta=90^{\circ}$, red symbols). 
 The out-of-plane field  $B_z=10$\,mT corresponds to the linear regime in Fig.~\ref{fig:firstfig}\textbf{c}. Empty and full symbols refer to two measurement sessions with higher resolution at low fields and lower resolution at high fields, respectively. We also report the results (grey symbols) of the control measurement performed on a sample with epitaxially grown Al on intrinsic GaAs. The solid red (blue) curve is obtained by fitting the experimental data for $\mathbf{B}_{ip}\perp \mathbf{I}$ ($\mathbf{B}_{ip}\parallel \mathbf{I}$) to the analytical model, see text. 
\textbf{b,} Polar plot showing the angle $\theta$ dependence of the vortex inductance for selected values of $B_z$.  
\textbf{c,} The color plot schematically represents the modulus of the order parameter, $|\Psi(x,y)|^2$, near the core of a pinned vortex, in the absence of in-plane field $B_{ip}$. The horizontal black arrow represents the direction of an oscillating current bias $\mathbf{I}\parallel \mathbf{\hat{x}}$, which exerts a Lorentz force $\mathbf{F} \parallel \mathbf{\hat{y}}$, white arrow. The measured vortex inductance is rotation symmetric and inversely proportional to the curvature of $|\Psi(x,y)|^2$ along $\mathbf{F}$. 
\textbf{d,} When a finite in-plane field is applied along $\mathbf{\hat{y}}$, the vortex core is squeezed as a consequence of the Rashba spin-orbit interaction. In this plot the curvature $k_x$ ($k_y$) along the $x$-axis ($y$-axis) is 1.64 (7.66) 
times that shown in panel \textbf{c}, reflecting the measured change in vortex inductance for $B_{ip}=1$\,T, see text. The curvature is always probed along $\mathbf{F}\parallel \mathbf{\hat{y}}$.
\textbf{e}-\textbf{f,} By rotating $\mathbf{B}_{ip}$ clockwise, the anisotropic vortex core rotates accordingly. 
Since the  $\mathbf{F}$ direction stays constant, such rotation makes it possible to probe the 
curvature of $|\Psi(x,y)|^2$ along an arbitrary direction and thus to extract its spatial tomography. The color range and level spacing is arbitrary, but the same in all the color plots.
} 
\label{fig:mainres}
\end{figure*}

Figure~\ref{fig:firstfig}\textbf{c} shows how the sample inductance depends on the out-of-plane magnetic field $B_z$. 
We notice that the function $L_v(B_z)$ is nearly linear up to 20\,mT (corresponding to $B_z \approx B_{c2}/3$), indicating that the inductance per added vortex is approximately constant. This means that the interaction between vortices is not relevant in this regime.
The measured ratio $L_v/B_z = 118$\,nH/mT is of the same order as the value expected from Eq.~(\ref{eq:vortind}) for a reasonable assumption of $k$~\cite{supplement}. 
At fields higher than 20\,mT, the $B_z$-dependence of the vortex inductance increases faster than linear (Fig.~\ref{fig:firstfig}\textbf{c}). In this regime, the order parameter in between the close-packed vortices is reduced compared to unperturbed value far from an isolated vortex. Hence, the curvature $k$ of $U(x,y)$ is reduced as well, leading to a super-linear dependence of $L_v$ on $B_z$.

More generally, any pair-breaking mechanism tends to reduce the curvature $k$.
As another example, Fig.~\ref{fig:firstfig}\textbf{e} shows the temperature dependence of the vortex inductance at $B_{z}=6.4$\,mT (linear low-field regime in Fig.~\ref{fig:firstfig}\textbf{c}). A pronounced increase of the vortex inductance is observed, which becomes very steep when the critical temperature is approached. Finally, pair-breaking by a purely in-plane magnetic field $B_{ip}$ must be reflected in the pure kinetic inductance $L_s$ too. This measurement is shown in   
Fig.~\ref{fig:firstfig}\textbf{f}. Notice the scale of the vertical axis: the kinetic inductance is 25 times smaller than the vortex inductance at 10\,mT, and it varies only by few nH for applied fields of the order of 1\,T~\footnote{It is worth to stress that, at least in our setup, the measurement of a pure kinetic inductance versus $B_z$ is difficult on macroscopic devices: on a millimeter scale, it is not possible to perfectly zero the out-of-plane field everywhere on the sample owing to the unavoidable inhomogeneity of the compensation field. The residual $B_z$ locally introduces vortices that have a sizable inductance which adds to the pure kinetic inductance.}. The sharp minimum for $|B_{ip}|<100$\,mT is likely due to the suppressed contribution of the Al wires used to bond the sample on the chip-carrier. Orbital pair-breaking is also seen in Fig.~\ref{fig:firstfig}\textbf{g}, which displays a reduction of $T_c(B_{ip})$ in $R(T,B_{ip})$ measurements. 

Having established the vortex inductance as a sensitive probe of the pinning strength $k$, we now come to our main observation, namely, an entirely unexpected \textit{increase} of the pinning strength controlled by the in-plane magnetic field. 
Figure~\ref{fig:mainres}\textbf{a} shows $L_v$ vs.~$B_{ip}$ at $B_z=10$\,mT (linear regime in Fig.~\ref{fig:firstfig}\textbf{c}), for both $\mathbf{B}_{ip}$ parallel (blue) and perpendicular (red)  to the direction of the drive current $\mathbf{I}\equiv dw\,\mathbf{j}$, where  the current density vector $\mathbf{j}$ is oriented along $\mathbf{\hat{x}}$, corresponding to the [110]-direction of InAs  (Fig.~\ref{fig:firstfig}\textbf{b}). 
In stark contrast to the behavior at $\mathbf{B}_z=0$ (Fig.~\ref{fig:firstfig}\textbf{f}),
a drastic and  surprising \textit{suppression} of the vortex inductance is seen for both  orientations when $B_{ip}$ is increased. At very high magnetic fields exceeding 2\,T, the inductance reaches a minimum and increases again near the in-plane critical field $B_{c,ip}\approx 2.7$~T, where it is expected to diverge. 
The full angle dependence of $L_v(\theta)$ is  displayed in Fig.~\ref{fig:mainres}\textbf{b} for $B_z=0$, 2, 5 and 10\,mT. The blue curve in Fig.~\ref{fig:mainres}\textbf{b} corresponds to the absence of vortices, i.e., the measured inductance is the kinetic inductance of the superfluid. The red curve in Fig.~\ref{fig:mainres}\textbf{b} corresponds to the same vortex density as in Fig.~\ref{fig:mainres}\textbf{a}. While the kinetic inductance is nearly isotropic at $B_{ip}=0$~\cite{supplement}, the vortex inductance shows a pronounced $\theta$-dependence with a two-fold symmetry. 
In order to check whether the effect results from SOI in the InAs quantum well, we have performed a control measurement on an Al film grown epitaxially on GaAs. There is no  2DEG  in GaAs and hence superconductivity is confined to the Al film. Moreover, in GaAs SOI is much smaller than in InAs, even when considering the effect of the Al/GaAs interface. For this device, the measured vortex inductance gradually increases with increasing in-plane field, see gray symbols in Fig.~\ref{fig:mainres}\textbf{a}. Importantly, this increase is almost perfectly isotropic~\cite{supplement}. 

Panels \textbf{c}-\textbf{f} of Fig.~\ref{fig:mainres} illustrate the order parameter profiles $|\Psi(x,y)|^2$ near the vortex cores as inferred from the measured reduction of the vortex inductance.   
Panel \textbf{c} shows $|\Psi(x,y)|^2 \propto U(x,y)$ in the vicinity ($x^2+y^2\ll \xi^2$) of the vortex center  for $\mathbf{B}_{ip}=0$. The vortex is assumed to be pinned at a point-defect located at the center of the figure. Since nothing breaks isotropy, the contours of constant $|\Psi(x,y)|^2$ are circles. 
As discussed above, the corresponding inductive voltage response reflects the curvature $k_y$ of  $|\Psi(x,y)|^2$ along $\mathbf{\hat{y}}$.

If an in-plane field is applied, e.g.~along $\mathbf{\hat{y}}$ ($\theta=90^{\circ}$, Fig.~\ref{fig:mainres}\textbf{d}), the vortex core will be squeezed in both the $\mathbf{\hat{x}}$ and the $\mathbf{\hat{y}}$ direction. However, the effect is more pronounced for the direction along $\mathbf{B}_{ip}$ (in this case $\mathbf{\hat{y}}$), i.e. $\partial_y^2 U(x,y)>\partial_x^2 U(x,y)$. Thus, for $\mathbf{B}_{ip}>0$ the contour lines of constant $|\Psi(x,y)|^2$ become ellipses with minor axis directed along $\mathbf{B}_{ip}$. By rotating $\mathbf{B}_{ip}$, the elliptic core will rotate accordingly. 
Such anisotropic vortex squeezing is the main result of our work. It is important to stress that it is the in-plane field $\mathbf{B}_{ip}$ that breaks the rotational symmetry. On the other hand, to \textit{detect} such anisotropy in the experiment, we use the supercurrent direction ($\mathbf{\hat{x}}\parallel \mathbf{I}$) as a reference.
Since in our device vortices oscillate along the $\mathbf{\hat{y}}\perp \mathbf{I}$ direction, the largest curvature $k_{\perp}$ (i.e., the smallest inductance) is probed for $\mathbf{B}_{ip}\perp \mathbf{I}$ ($\theta=90^{\circ}$, Fig.~\ref{fig:mainres}\textbf{d})  while the smallest curvature $k_{\parallel}$ (largest inductance) is probed for  $\mathbf{B}_{ip}\parallel \mathbf{I}$ ($\theta=0^{\circ}$, Fig.~\ref{fig:mainres}\textbf{f}). As $\mathbf{B}_{ip}$ is continuously rotated, inductance measurements provide a tomography of the order parameter in the vicinity of the vortex center, as shown in Fig.~\ref{fig:mainres}\textbf{b}.
The effect is remarkably robust: for $B_{ip}=1$\,T, $k_{\perp}$ ($k_{\parallel}$) increases by a factor 7.66 (1.64) compared to the $B_{ip}=0$ case, as deduced from the corresponding reduction of $L_v$ in  Fig.~\ref{fig:mainres}\textbf{a}, red (blue) curve.

Now we turn to possible explanations for the striking observations in Figure~\ref{fig:mainres}.
The key findings, that must be captured by a theoretical model, are: 
(i) the vortex inductance anomalously decreases with the applied in-plane field;
(ii) the decrease is anisotropic, with a two-fold symmetry,  (iii) it is maximal (minimal) when the field is perpendicular (parallel) to the current density; 
(iv) the effect is visible only in epitaxial Al/InAs 2DEG devices, while it is absent in the control Al/GaAs samples without 2DEG and with largely reduced SOI. 

%
 The non-centrosymmetry of the quasi-2D film is captured on the microscopic level by the isotropic Rashba Hamiltonian $H_R = \alpha_R (\mathbf{k}\times\mathbf{n})\cdot\boldsymbol{\sigma}$, where the unit vector $\mathbf{n}$ (along the polar axis) is normal to the plane of the superconducting film. We estimate the Rashba coupling $\alpha_R$ and the $g$-factor in the Zeeman Hamiltonian $H_Z = g\mu_B \mathbf{B}\cdot\boldsymbol{\sigma}$ to be of the  order of 
$\alpha_R=15$\,meV$\cdot$nm and $g=-10$, respectively.

On the other hand, as shown by Edelstein~\cite{Edelstein1996}, the joint effect of the Rashba SOI, in-plane magnetic field and superconducting pairing can be described, within the Ginzburg-Landau approach, by adding a new term to the free energy---the so-called Lifshitz invariant.
As discussed below, it is the Lifshitz term which can explain the
anisotropic vortex squeezing, in combination with the in-plane field. 
The Ginzburg-Landau free energy density in question has the following form:
\begin{equation}
    F[\Psi,\mathbf{A}]=a(T)|\Psi|^2+\frac{b}{2}|\Psi|^4+\frac{|\mathbf{D}\Psi|^2}{4m}+\frac{\mathbf{B}^2}{2\mu_0}+F_L[\Psi,\mathbf{A}],
\end{equation}
where the last two terms correspond to the magnetic energy density and to the (isotropic) \textit{Lifshitz invariant}~\cite{Edelstein1996}:
\begin{align}\label{eq:IsotropicLifshitz}
    F_L[\Psi,\mathbf{A}] 
    &= -\frac{1}{2}\kappa (\mathbf{n}\times\mathbf{B})\cdot\bigl[(\Psi)^*\mathbf{D}\Psi+\Psi(\mathbf{D}\Psi)^*\bigr].
\end{align}
The Lifshitz invariant is the direct manifestation of Rashba  SOI at the Ginzburg-Landau level. 
In the above expressions $\Psi$ stands for the condensate wave function, $\mathbf{A}$ for the vector potential, $\mathbf{B}=\text{rot}\,\mathbf{A}$ for the corresponding (in-plane + out-of-plane) magnetic field, and
$\mathbf{D}=\tfrac{\hbar}{i}\mathbf{\nabla}-2e\mathbf{A}$ for the covariant momentum operator ($|e|$ is the 
elementary charge). 
Here and below we assume Rashba SOI with (at least) $C_{4v}$ point-group symmetry in the sample plane. The whole symmetry of $F[\Psi,\mathbf{A}]$ is, however, lowered to $C_{2v}$ when the in-plane magnetic field and current drive 
are present. 

On the phenomenological level, a figure of merit 
quantifying the impact of the non-centrosymmetry is given by the parameter 
$\kappa$ or the Lifshitz-Edelstein length $\ell_\kappa$:
\begin{align}
    \kappa &\simeq 3\frac{\alpha_R}{\hbar} \frac{g\mu_{B}}{v_F p_F}, \ \ \ \ \ \ell_\kappa=\frac{b}{2|\kappa|\mu_0 |e| a(T)},
\end{align}
where $\mu_B$ is the Bohr magneton, and $p_F$ and $v_F$ stand for the Fermi momentum and velocity (for a complete derivation, including the numerical prefactor, see Ref.~\cite{Edelstein1996}).
Upon functional variation of the extended Ginzburg-Landau free energy density $F[\Psi,\mathbf{A}]$, one obtains the first and second Ginzburg-Landau equation for 2D Rashba-superconductor, as  discussed in the Supplementary Material~\cite{supplement}.

The goal of our analytical calculation is to describe the 
wave function of the vortex order parameter, $\Psi_v(x,y)$, in the vicinity of the vortex core at $(x_0=0,y_0=0)$, therefore we assume the following asymptotic form 
\begin{equation}\label{Eq:PSIwf}
    \Psi_v(x,y)= K\cdot(x+i\delta\, y)\,\exp{\left[\frac{p}{2}\,x^2 + q\, x y + \frac{r}{2}\,y^2\right]},
\end{equation}
where the real parameters $K$, $\delta$, $p$, $q$ and $r$ can be determined~\cite{supplement} from the 
Ginzburg-Landau equations including the Lifshitz term~\footnote{The ansatz for $\Psi_v(x,y)$ has the same asymptotic form (including the cubic terms for the vortex core region, 
i.e.~for $x/\xi, y/\xi\ll 1$) as the Abrikosov solution in the conventional case \cite{AbrikosovJETP}.}.  
Obviously, for $\delta\neq 1$ the vortex factor $K\cdot(x+i\delta\, y)$ of the solution $\Psi_v$(x,y) possesses a certain ellipticity mimicking the reduction of symmetry to $C_{2v}$ due to an in-plane magnetic field.
When discussing the model, it is customary to assume a fixed direction of $\mathbf{B}_{ip}\parallel\mathbf{\hat{y}}$,
while in the experiment it is the direction of current that is kept fixed $\mathbf{I}\parallel\mathbf{\hat{x}}$. In the limit of a point-like pinning defect, the effective vortex potential $U(x,y)$ mirrors $|\Psi_v(x,y)|^2$, 
hence
\begin{equation}
U(x,y)\equiv \tfrac{1}{2}k_{x}\, x^2 + \tfrac{1}{2}k_{y}\, y^2\simeq |\Psi_v(x,y)|^2\simeq K^2\,x^2 +K^2\delta^2\, y^2. 
\end{equation}
%
The theoretical values of the curvatures $k_x=\partial_x^2U(0,0)\simeq2K^2$ and $k_y=\partial_y^2U(0,0)\simeq2K^2\delta^2$ of $U(x,y)$ can be directly linked to the experimentally determined curvatures (inductances) $k_{\parallel}\  (L_{v,\parallel})$ and $k_{\perp}\ (L_{v,\perp})$, where the subscripts $\parallel$ and $\perp$ discriminate correspondingly between the mutual orientations of $\mathbf{B}_{ip}$ and $\mathbf{I}$ in the experiment, particularly,
$k_{\parallel}=k_x$ and $k_{\perp}=k_y$. 
The independently measured input parameters for our model are $\xi=73$\,nm, $\lambda=227$\,nm and $B_z=10$~mT.
Theory provides a set of algebraic equations 
for $k_{x}$ and $k_{y}$ as functions of $B_{ip}$. The equations contain the Lifshitz-Edelstein length $\ell_{\kappa}$ and the effective thermodynamic critical field $B_c^{\ast}$ as parameters that can be determined by fitting data in Fig.~\ref{fig:mainres}\textbf{a}, using Eq.~(\ref{eq:vortind}) to link curvature to inductance.
Restricting the fit to the range of [-0.1\,T, 0.1\,T] we obtain~\cite{supplement} $B_c^{\ast} = 96$\,mT and $\ell_\kappa = 590$\,nm. The resulting fitting curves are shown as solid lines in  Fig.~\ref{fig:mainres}\textbf{a}. 
Despite the simplified phenomenological approach, our model quantitatively captures (i) the increase of \textit{both} curvatures, $k_{\parallel}$ and $k_{\perp}$, upon application of an in-plane field, as well as (ii) the anisotropy ratio $k_\perp/k_\parallel>1$ of the two curvatures. 
For $B_{ip}>0.1\,$T the fits underestimate $L_v$, most probably  because the quadratic approximation of $\Psi(\mathbf{r})$ at the vortex cores is no 
longer valid.

In order to further substantiate our interpretation of the reduced vortex inductance as an enhanced pinning strength, we investigate an entirely different signature of pinning, i.e., the depinning critical current. 
If the local minima of $U(\mathbf{r})$  become sharper in in-plane field, then one would expect that not only its bottom curvature will increase, but also its maximal slope, i.e.~$\max [|\partial_{\mathbf{r}} U(\mathbf{r})|]$. 
This corresponds to the maximal restoring force that pinning centers can exert before the depinning point. On the basis of the vortex inductance measurements just discussed, the depinning current is expected to display a similar peculiar increase with the in-plane field. 
We performed such measurements on a separate sample from the same wafer that was designed in a standard Hall bar geometry. The width was reduced to 2.3\,\textmu m, leading to a smaller critical current and thus less Joule heating.

Owing to the large contact resistance of this particular sample, a large heat is generated at the bonding pads of the device which reaches the device through the substrate in a fraction of a second. Therefore, the IV traces were acquired in 9~ms, with 30~s waiting time needed to cool the system back to the base temperature $T=0.1$~K. The sweep time was chosen in such a way that a further reduction of the sweep time did not affect the depinning current anymore. Each measurement point corresponds to the average of the depinning currents resulting from 45 repetitions of the IVs.

\begin{figure}[t]
	\includegraphics[width=\columnwidth]{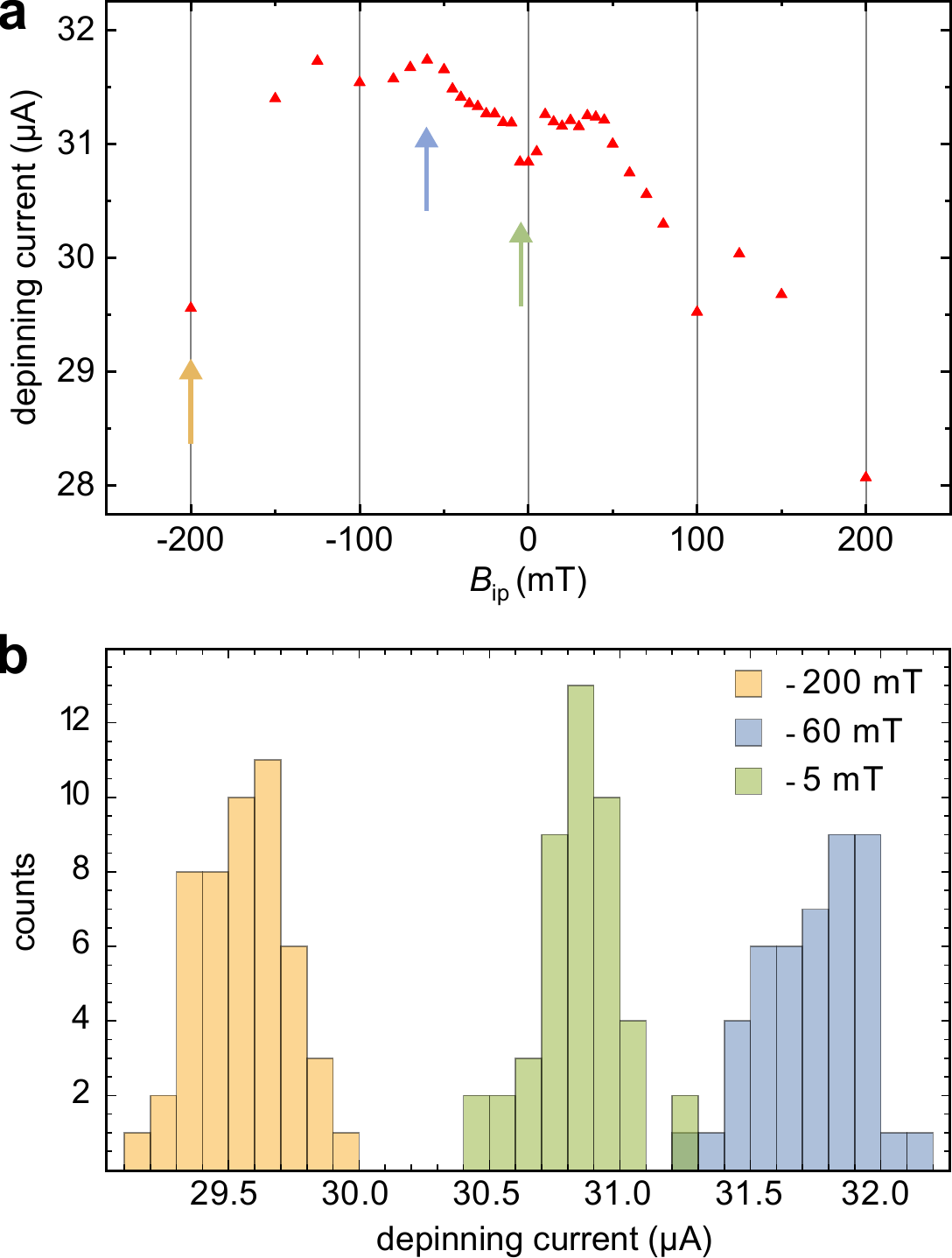}
	\caption{\textbf{DC transport signature of the field-enhanced pinning.} \textbf{a,} Depinning current as a function of the in-plane magnetic field component perpendicular to the current ($B_y$), measured for $B_z=5$~mT. Each data point is obtained from the average of 45 repeated IV measurements.
	\textbf{b,} Histograms showing the distribution of the depinning current values for three selected values of $B_y$, namely, -5~mT (light green), -60 mT (light blue) and -200~mT (light orange). These values are indicated by arrows of the corresponding colors in the top panel.
	}
	\label{fig:depin}
\end{figure}

The results of these depinning current measurements for $B_z=5$~mT are shown in Fig.~\ref{fig:depin}\textbf{a}. To maximize the effect, the in-plane field is oriented perpendicular to the current. ($\mathbf{B}_{ip}=B_y\mathbf{\hat{y}}$, $\theta=90^{\circ}$).  We observe a minimum for the depinning current at zero in-plane field, a maximum at about $|B_{ip}|\approx 60$~mT and then eventually a rapid suppression for $|B_{ip}| >100$~mT, when pair-breaking becomes significant.  Figure~\ref{fig:depin}\textbf{b} shows selected histograms for the depinning current distribution corresponding to selected values of $B_y$. The distributions are relatively narrow: the data scatter is mostly due to the extreme sensitivity of the measurement to the precise value of the effective $B_z$, which we kept constant at 5~mT using a field compensation routine, discussed in the Supplementary Material~\cite{supplement}.
Further measurements on the same and on another device are discussed in the Supplementary Material~\cite{supplement}. These results confirm the outcome of the vortex inductance measurements, i.e., the pinning interaction is anisotropically enhanced by a moderate magnetic field. 
Hence, the depinning current provides a further independent evidence of the apparently anomalous shrinking of the vortex cores.

Before concluding, we want to briefly discuss alternative (or additional) mechanisms that might be responsible for the 
anisotropic vortex inductance in 2D Rashba-superconductors. 
The simplest possible origin of anisotropy in $U(\mathbf{r})$ is through a magnetic field-induced anisotropy in the coherence length $\xi$, e.g., by an anisotropic Fermi velocity. The Rashba SOI spin-splits the Fermi surface while preserving its circular symmetry. An applied in-plane magnetic field shifts and distorts the two circular Fermi surfaces. However, for a realistic Rashba coefficient $\alpha_R=15$\,meV$\cdot$nm~\cite{Baumgartner2021}, we obtain a relative anisotropy in $v_F$ of the order of just $10^{-3}$, too small to explain the marked anisotropy observed in the experiment. Moreover, the measured anisotropy of the in-plane critical field~\cite{supplement} is also much smaller than that of $L_v$. 

Another interesting possibility is that in an InAs 2DEG proximitized by an epitaxial Al layer the pairing function is not purely of $s$-wave type, but rather admixed of $p_x+ip_y$-wave. Without an in-plane field, the modulus of the pairing function is still isotropic both in the reciprocal and in the real space. The application of an in-plane field gradually projects the $p_x+ip_y$-wave pairing into its $p_y$-wave component. The anisotropic $\Delta(\mathbf{k})$ produces, upon Fourier transform, an anisotropic coherence length $\xi$. This argument does not explain, \textit{per se}, the increase in the pinning force with the magnetic field. However, Hayashi and Kato~\cite{HKPRB2002,KHPhysC2003} have found that for sharp defects (in the sense discussed in  Ref.~\cite{ThunebergPRB84}) and for $p_x+ip_y$-wave pairing, the pinning potential becomes steeper near the vortex core center.
Nevertheless, this picture cannot adequately explain the rapid decrease of the vortex inductance with small or moderate in-plane fields.  
We  therefore believe that our model based on the Lifshitz invariant provides a more natural explanation of both the pinning enhancement and the field induced anisotropy. 
On the other hand, our work is compatible with the occurrence of a $p_x+ip_y$-wave pairing. The  latter is at the basis of many proposals aiming at implementing topological superconductivity in InAs 2DEG proximitized by epitaxial Al. Recent experiments indicate that unconventional pairing affects the kinetic inductance in such systems~\cite{Phan2021}. Our data evidence the caveat that, in the presence of residual vortices, the inductance resulting from their oscillations around pinning centers can dominate over any other inductance contribution.


In conclusion, we have demonstrated experimentally that in 2D Rashba-superconductors the application of an in-plane field squeezes the vortex cores, leading to an enhanced pinning. The vortex squeezing is anisotropic---vortex cores are more compressed if the in-plane magnetic field is orientated perpendicular to the supercurrent. 
By rotating the in-plane field, the inductance measurements provide a tomography of order parameter profile near the vortex core.
These results constitute a clear manifestation of Lifshitz invariants in a macroscopic observable of a synthetic Rashba superconductor. Similar to Skyrmions and chiral solitons \cite{fert_skyrmions_2017,bogdanov_NatRev_2020,GOBEL2021} in magnetic systems with Lifshitz invariants, also Abrikosov vortices are strongly affected by the broken symmetries. Moreover, our study opens the path towards the manipulation of quasiparticles in the vortex cores~\cite{Caroli1964} including Majorana fermions \cite{Xu_JinFengJia2015}.

\section{Acknowledgments} 
We thank A.~Drachmann and C.~Marcus for providing chips with Al/GaAs reference films and T.~Lemberger, J.~Schmalian, E.~Bauer, M.~Sigrist, D.~Agterberg, Y.~Ando, L.~Tosi and C.~Timm for helpful discussions. Work at Regensburg University was funded by the Deutsche Forschungsgemeinschaft (DFG, German Research Foundation) – Project-ID 314695032 – SFB 1277 (Subprojects B05, B07, and B08). D.K.~acknowledges a partial support from the project 
IM-2021-26 (SUPERSPIN) funded by Slovak Academy of Sciences via the programme IMPULZ 2021.


\section{Competing interests}
The authors declare  no competing interests.

\section{Additional Information}
\textbf{Supplementary Materials:} 
Theory: Ginzburg-Landau (GL) Equations in the presence of the isotropic Lifshitz invariant; First GL equation for a thin superconducting film; Experiment: Experimental methods; Isotropy of the kinetic inductance in the absence of vortices; Isotropy of kinetic inductance and vortex inductance in Al/GaAs samples;  Weak anisotropy for $B_{c,ip}$; Vortex inductance for low in-plane fields; Further measurements of the depinning current; Out-of-plane field compensation procedure.

\clearpage
\newpage

\onecolumngrid
\begin{center}
\textbf{\Large Supplementary Information} 

\vspace{0.5cm}

\noindent \textbf{\large Anisotropic vortex squeezing in synthetic Rashba superconductors: \newline a manifestation of Lifshitz invariants}
\end{center}

\beginsupplement

\section{Ginzburg-Landau Equations in the presence of the isotropic Lifshitz invariant: general considerations}




On a phenomenological level, non-centrosymmetric polar superconductors governed by the Rashba spin-orbit coupling,  
and subjected to an external magnetic field $\mathbf{B}=\text{rot}\,\mathbf{A}$, can be described by the Ginzburg-Landau (GL) functional for the free energy density \cite{Edelstein1996}:
\begin{equation}
    F[\psi,\mathbf{A}]=a(T)|\psi|^2+\frac{b}{2}|\psi|^4+\frac{1}{4m}(\mathbf{D}\psi)^*\cdot\mathbf{D}\psi
    +\frac{\mathbf{B}^2}{2\mu_0}+F_{L}[\psi,\mathbf{A}]
    \equiv 
    F_{GL}[\psi,\mathbf{A}]+F_L[\psi,\mathbf{A}],
\end{equation}
where the pre-last term represents density of the magnetic energy and the last one the so called isotropic 
\textit{Lifshitz invariant}.
The latter stems from an interplay of the Rashba spin-orbit interaction, 
$H_R = \alpha_R(\mathbf{k}\times\mathbf{n})\cdot\boldsymbol{\sigma}$, Zeeman coupling, $H_Z = g\mu_B \mathbf{B}\cdot\boldsymbol{\sigma}$ and the superconducting coherence, and its explicit form reads
\cite{Edelstein1996}:
\begin{align}
    F_L[\psi,\mathbf{A}] &= -\frac{1}{2}\kappa (\mathbf{n}\times\mathbf{B})\cdot \mathbf{Y}_\psi
    \equiv -\frac{1}{2}\kappa (\mathbf{n}\times\mathbf{B})\cdot\bigl[(\psi)^*\mathbf{D}\psi+\psi(\mathbf{D}\psi)^*\bigr],
\end{align}
wherein the Edelstein parameter $\kappa$,
\begin{align}
    \kappa &= 3\frac{\alpha_R}{\hbar}\frac{g\,\mu_{B}}{v_F\,p_F}\,f_3\left(\frac{\alpha_R\,p_F}{\hbar\,\pi\, k_{\mathrm{B}} T_c}\right)\simeq 3\frac{\alpha_R}{\hbar}\frac{g\,\mu_{B}}{v_F\,p_F}
    \ \ \text{with}\ \ f_3(x)\simeq 0.475\int\limits_{0}^{\pi} \mathrm{d}t \sum\limits_{n=0}^{\infty}\frac{\sin{t}\,(x\sin{t})^2}{(2n+1)^3[(2n+1)^2+(x\sin{t})^2]},
\end{align}
serves as a figure of merit of the non-centrosymmetry of polar systems. 
The meaning of different symbols/letters is set by the following notation, moreover, all quantities are assumed 
to be in SI units:
\begin{itemize}
    \item $\psi$ stands for the condensate (Cooper pair) macroscopic wave function in the GL approach, being in $D$ spatial dimensions, the unit of $\psi$ is $\text{m}^{-D/2}$,
    \item $\mathbf{n}$ is the spin-orbit-coupling unit vector defining the polar axis, without a loss of generality we assume $\mathbf{n}=\hat{z}=(0,0,1)$, and $\alpha_R$ is the corresponding Rashba coupling,
    \item $\boldsymbol{\sigma}$ stands for the vector $\{\sigma_x,\sigma_y,\sigma_z\}$ of Pauli spin $\tfrac{1}{2}$-matrices, spin quantization axis is along $\mathbf{n}$,
    \item $\mathbf{A}$ is the vector potential and $\mathbf{B}=\text{rot}\,\mathbf{A}$ is the total magnetic field (in-plane + out-of-plane w.r.t.~the given polar axis $\mathbf{n}$). We employ the Coulomb gauge, i.e., $\text{div}\,\mathbf{A}=0$,
    \item $\mathbf{D}=\tfrac{\hbar}{i}\mathbf{\nabla}-2e\mathbf{A}$ is the covariant momentum operator,
    \item $m$, $g$ and $e<0$ are, correspondingly, the electron effective mass, effective g-factor and the electron charge,
    \item $\mu_0$ and $\mu_B$ are, correspondingly, magnetic permeability and the Bohr magneton,
    \item $a(T)=\alpha_0 (T-T_c)/T_c\leq 0$ and $b$ are the conventional GL parameters, $T_c$ is the critical temperature, and $k_{\mathrm{B}}$ the Boltzmann constant,
    \item $p_F$ and $v_F$ stand for the Fermi momentum and Fermi velocity,
    \item $\kappa$ is the Edelstein coefficient of non-centrosymmetry (not to be confused with the Ginzburg-Landau parameter $\kappa_{GL}\equiv \lambda/\xi$), in SI units $\kappa$ is expressed in  m$\cdot$C/kg. 
\end{itemize}

Considering $F[\psi,\mathbf{A}]$ as a functional of $\psi$ and $\mathbf{A}$, 
one derives in a standard way the first and second GL equations in the presence of non-centrosymmetry, along with a boundary condition on the interface between the superconductor and a vacuum (an insulator or normal metal):
\begin{align}
\text{1st GL-Eq:\ } & 0=\frac{\delta F}{\delta \psi^*} &   
0 &=\frac{1}{4m}\bigl[\mathbf{D}-2m\kappa (\mathbf{n}\times\mathbf{B})\bigr]^2\psi+\bigl[a(T)-m\kappa^2(\mathbf{n}\times\mathbf{B})^2\bigr]\psi 
      + b|\psi|^2\psi,\label{Eq:GL1}
\\ 
\text{2nd GL-Eq:\ } & 0=\frac{\delta F}{\delta \mathbf{A}} & 
0 &= \text{rot}\,\left[\frac{1}{\mu_0}\mathbf{B}+\frac{1}{2}\kappa\, (\mathbf{n}\times\mathbf{Y}_\psi)\right]-\frac{e}{2m}\mathbf{Y}_\psi+2\kappa e |\psi|^2 (\mathbf{n}\times\mathbf{B}), \label{Eq:GL2}  
\\
\text{boundary condition:\ } &{}  & 
0 &= \boldsymbol{\hat{\nu}}_{\text{out}}\cdot\bigl[\mathbf{D}-2m\kappa (\mathbf{n}\times\mathbf{B})\bigr]\psi\Bigl|_{\text{interface}} \label{Eq:GL3},  
\end{align}
where $\boldsymbol{\hat{\nu}}_{\text{out}}\ (\text{in general not necessarily related with}\ \mathbf{n})$ is the outer normal vector pointing from the superconductor to the vacuum (an insulator or normal metal).
For a completeness, the 2nd GL equation can be written in the form of a Maxwell equation---namely the Ampere law:
\begin{align}
     \text{rot}\,\boldsymbol{\mathcal{H}}  &=\mathbf{j}_s\,, \ \ \ \ \ \text{where}\ \ \ \ \    
     &   \text{magnetic field intensity:}\ \ \ \ \boldsymbol{\mathcal{H}} & = \frac{1}{\mu_0}\mathbf{B} - \boldsymbol{\mathcal{M}},\\
                                         &{}              
     &   \text{magnetization:}\ \ \ \boldsymbol{\mathcal{M}} &= -\frac{1}{2}\kappa\, (\mathbf{n}\times\mathbf{Y}_\psi),\\
                                         &{}              
     &   \text{supercurrent:}\ \ \ \ \, \mathbf{j}_s &= \frac{e}{2m}\mathbf{Y}_\psi-2\kappa e |\psi|^2 (\mathbf{n}\times\mathbf{B}).
\end{align}

\section{1st GL equation for a thin superconducting film in  in-plane and out-of-plane magnetic field}

In what follows we assume that:
\begin{itemize}
\item the superconducting film is quasi-2D, lies in $xy$-plane, and is terminated on both sides by a vacuum~i.e.~$\pm\boldsymbol{\hat{\nu}}_{\text{out}}\parallel\mathbf{n}=\hat{z}=(0,0,1)$, 
\item the film has an \emph{extrapolation length} $d$ and extends in the transverse direction within the interval $z\in[-\tfrac{d}{2},+\tfrac{d}{2}]$; the extrapolation length  $d= d_{\text{geo}}+\mathcal{O}(1)\xi$, where $d_{\text{geo}}$ is the true geometrical (i.e., physical) thickness of the Al film and $\xi$ is the GL coherence length,
\item the film is without pronounced in-plane crystallographic anisotropies, i.e., we assume in-plane $C_{4v}$ (or higher $C$) symmetry---this is imprinted in the form of $H_R$,
\item the vortex-generating (out-of-plane) component of magnetic field, $\mathbf{B}_z=B_z\hat{z}=(0,0,B_z)$, is perpendicular to the film,
\item the in-plane component of magnetic field $\mathbf{B}_{ip}$ is pointing along the $y$-axis,~i.e.~$\mathbf{B}_{ip}=B_y\hat{y}=(0,B_y,0)$,
\end{itemize}
then
\begin{itemize}
\item then the total magnetic field vector $\mathbf{B}=\mathbf{B}_{z}+\mathbf{B}_{ip}=(0,B_y,B_z)$, 
and the corresponding vector potential in the Coulomb gauge 
$\mathbf{A}=(-B_z\tfrac{y}{2}, B_z\tfrac{x}{2}, -B_y x)$,
\item $\mathbf{n}\times\mathbf{B}=-B_y\hat{x}=(-B_y,0,0)$ $\Rightarrow$ 
$m\kappa^2 (\mathbf{n}\times\mathbf{B})^2=m\kappa^2 B_y^2$.
\end{itemize}
As a comment, $B_z$ and $B_y$ are local magnetic fields near the vortex core and in principle they differ from the corresponding laboratory values. However, we expect the difference to be small owing to the fact that the film thickness is much smaller than both $\xi$ and $\lambda$. Thus, the in-plane field is not expected to be significantly affected by the negligible in-plane screening currents. However, in this respect, it is difficult to model the role of the 2DEG, which is relatively thicker compared to the Al film.
Also the out-of-plane component is expected to be similar to the applied $B_z$ field, since the Pearl length $2\lambda^2/d$ is of the order of many micrometers.\\

\noindent We will solve the 1st GL equation near the vortex core region, for which we assume that $B_y$ and $B_z$ are not varying in space on the length scale of the coherence length $\xi$, and therefore we treat them as constants. 
Let us elaborate in detail on the 1st GL equation, Eq.~(\ref{Eq:GL1}):
\begin{equation}
0 =\frac{1}{4m}\bigl[\mathbf{D}-2m\kappa (\mathbf{n}\times\mathbf{B})\bigr]^2\psi+\bigl[a(T)-m\kappa^2(\mathbf{n}\times\mathbf{B})^2\bigr]\psi 
      + b|\psi|^2\psi\,.
      \label{eq:origvar}
\end{equation}
We consider each term individually.
\begin{itemize}
    \item $[\mathbf{D}-2m\kappa (\mathbf{n}\times\mathbf{B})]=\frac{\hbar}{i}\boldsymbol{\nabla}+(e B_z y+2m\kappa B_y, -e B_z x, 2e B_y x)$,
    \item $[\mathbf{D}-2m\kappa (\mathbf{n}\times\mathbf{B})]^2
    =-\hbar^2\Delta+2\frac{\hbar}{i}(e B_z y+2m\kappa B_y, -e B_z x, 2e B_y x)\cdot\boldsymbol{\nabla}+[(e B_z y+2m\kappa B_y)^2+e^2(B_z^2 + 4 B_y^2)x^2]$,
    \item shifting the $y$-coordinate $y = y_{new}-2\kappa m B_y/(eB_z)$ $\Rightarrow$
    \begin{align}
    [\mathbf{D}-2m\kappa (\mathbf{n}\times\mathbf{B})]^{\phantom{2}}&=\frac{\hbar}{i}\boldsymbol{\nabla}+(e B_z y_{new}, -e B_z x, 2e B_y x)\\
    &=[\text{renaming:}\ y_{new}\mapsto y]\nonumber\\
    &=\frac{\hbar}{i}\boldsymbol{\nabla}+(e B_z y, -e B_z x, 2e B_y x)\\
    \nonumber\\
    [\mathbf{D}-2m\kappa (\mathbf{n}\times\mathbf{B})]^2&=
    -\hbar^2\Delta
    +2\tfrac{\hbar}{i}(e B_z y_{new}, -e B_z x, 2e B_y x)\cdot\boldsymbol{\nabla}+[e^2 B_z^2 y_{new}^2+e^2(B_z^2 + 4 B_y^2)x^2]\\
    &=[\text{renaming:}\ y_{new}\mapsto y]\nonumber\\
    &=-\hbar^2\Delta
    +2\tfrac{\hbar}{i}(e B_z y, -e B_z x, 2e B_y x)\cdot\boldsymbol{\nabla}+[e^2 B_z^2 y^2+e^2(B_z^2 + 4 B_y^2)x^2]\\
    &=-\hbar^2\Delta
    -2eB_z \hat{L}_z+4 e B_y \left(x \tfrac{\hbar}{i}\partial_z\right)+[e^2B_z^2y^2+e^2(B_z^2 + 4 B_y^2)x^2]
    \end{align}
    \item $\bigl[a(T)-m\kappa^2(\mathbf{n}\times\mathbf{B})^2\bigr]=a(T)-m\kappa^2 B_y^2=-|a(T)|-m\kappa^2 B_y^2\leq 0$.
    \end{itemize}
The shift of the $y$ coordinate is mathematically immaterial, since it can be absorbed into a change of the origin of $y$-axis and in a redefinition of the order parameter wave function: 
$$
\psi(x,y,z)=\psi(x,y_{new}-\tfrac{2\kappa m B_y}{eB_z},z)\equiv\Tilde{\psi}(x,y_{new},z).
$$
For simplicity of notation and minimal proliferation of symbols, we denote $y_{new}$ by $y$ and $\Tilde{\psi}(x,y,z)$ by $\psi(x,y,z)$. On the physical ground $\Delta y =\tfrac{2\kappa m B_y}{eB_z}$ gives a displacement of the vortex core 
centre---defined as a minimum of $|\psi|$---from the position of a maximum of the out-of-plane magnetic field penetrating the vortex.\\

\noindent After shifting and renaming the equation~(\ref{eq:origvar}) reads:
\begin{equation}
    0 = -\tfrac{\hbar^2}{4m}\Delta\psi
    -\tfrac{2eB_z}{4m} \hat{L}_z\psi+\tfrac{4 e B_y}{4m} \left(x \tfrac{\hbar}{i}\partial_z\right)\psi+\tfrac{e^2}{4m}[B_z^2y^2+(B_z^2 + 4 B_y^2)x^2]\psi-[|a(T)|+m\kappa^2 B_y^2]\psi+b|\psi|^2\psi.
\end{equation}

\noindent We look for a solution $\psi$ that separates the in-plane dependence from the transverse one:
\begin{equation}
    \psi(x,y,z)\equiv\Psi(x,y)\cdot\Phi(z),
\end{equation}
where $\Psi(x,y)$ accounts for the in-plane and $\Phi(z)$ for the out-of-plane order parameter wave function, respectively. 
Assuming $\Phi(z)$ is a real-valued function (transverse domain is simply connected), then
\begin{align}
0 &= -\tfrac{\hbar^2}{4m}\Phi\cdot(\partial_{xx}+\partial_{yy})\Psi-\tfrac{\hbar^2}{4m}\Psi\cdot\partial_{zz}\Phi
    -\tfrac{2eB_z}{4m}\Phi\cdot \hat{L}_z\Psi
    +\tfrac{4 e B_y}{4m}\Psi\cdot \left(x \tfrac{\hbar}{i}\partial_z\right)\Phi\nonumber\\
    &{}\nonumber\\
&+\tfrac{e^2}{4m}[B_z^2y^2+(B_z^2 + 4 B_y^2)x^2]\Psi\cdot\Phi-[|a(T)|+m\kappa^2 B_y^2]\Psi\cdot\Phi+b|\Psi|^2\Psi\cdot|\Phi|^2\Phi.
\end{align}
Now, we average the last equation along the transverse direction within the extrapolation length 
$d= d_{\text{geo}}+\mathcal{O}(1)\xi$: 
\begin{align}
0 &= -\tfrac{\hbar^2}{4m}\langle\Phi\rangle\cdot(\partial_{xx}+\partial_{yy})\Psi
-\tfrac{\hbar^2}{4m}\Psi\cdot\langle\partial_{zz}\Phi\rangle
    -\tfrac{2eB_z}{4m}\langle\Phi\rangle\cdot \hat{L}_z\Psi
    +\tfrac{4 e B_y}{4m}\Psi\cdot \langle\left(x \tfrac{\hbar}{i}\partial_z\right)\Phi\rangle\nonumber\\
    &{}\nonumber\\
&+\tfrac{e^2}{4m}[B_z^2y^2+(B_z^2 + 4 B_y^2)x^2]\Psi\cdot\langle\Phi\rangle-[|a(T)|+m\kappa^2 B_y^2]\Psi\cdot\langle\Phi\rangle+b|\Psi|^2\Psi\cdot\langle|\Phi|^2\Phi\rangle,\label{Eq:GL1_separ}
\end{align}
where the meaning of the angular brackets (transverse averaging) is as follows:
\begin{equation}
  \langle f=f(z)\rangle=\frac{1}{d}\int_{-\tfrac{d}{2}}^{+\tfrac{d}{2}}dz\ f(z)\,.
\end{equation}
Doing so, we obtain:
\begin{align}
    \langle\partial_{zz}\Phi\rangle &=\tfrac{1}{d}\Bigl[\partial_z\Phi\bigl|_{z=+\tfrac{d}{2}}\ -\ \partial_z\Phi\bigl|_{z=-\tfrac{d}{2}}\Bigr]=0,\\
    \langle\partial_{z}\Phi\rangle  &=\tfrac{1}{d}\Bigl[\Phi(z=+\tfrac{d}{2})\ -\ \Phi(z=-\tfrac{d}{2})\Bigr]=0,\\
    \langle\Phi\rangle &= \text{real constant}\neq 0.
\end{align}
In the equations above, we set to zero $\langle\partial_{zz}\Phi\rangle$ and $\langle\partial_{z}\Phi\rangle$. This is due to the boundary conditions, Eq.~(\ref{Eq:GL3}), that should be satisfied at two interfaces located at $z=\pm\tfrac{d}{2}$ ($\boldsymbol{\nu}_{\text{out}}=\pm\hat{z}$), we treat both in conjunction:
    \begin{align}
        0 &= \boldsymbol{\nu}_{\text{out}}\cdot\bigl[\mathbf{D}-2m\kappa (\mathbf{n}\times\mathbf{B})\bigr]\Psi(x,y)\cdot\Phi(z)\bigl|_{z=\pm\tfrac{d}{2}}
        =\bigl[\boldsymbol{\nu}_{\text{out}}\cdot\mathbf{D}-2m\kappa\, \underbrace{\boldsymbol{\nu}_{\text{out}}\cdot(\mathbf{n}\times\mathbf{B})}_{=0}\bigr]\Psi(x,y)\cdot\Phi(z)\bigl|_{z=\pm\tfrac{d}{2}}
        \\
        &\Updownarrow\nonumber\\
        \nonumber\\
        0 &= \tfrac{\hbar}{i}\Psi(x,y)\cdot\partial_z\Phi(z)\bigl|_{z=\pm\tfrac{d}{2}}\ \pm\ 2e B_y x\, \Psi(x,y)\cdot\Phi(z=\pm\tfrac{d}{2}),\\
        \nonumber\\
        &\Updownarrow \text{assuming}\ \Psi(x,y)\neq 0\nonumber\\
        \nonumber\\
        0 &= \tfrac{\hbar}{i}\partial_z\Phi(z)\bigl|_{z=\pm\tfrac{d}{2}}\ \pm\ 2e B_y x\,\Phi(z=\pm\tfrac{d}{2}).
    \end{align}
The above two equations should be satisfied, correspondingly, for any point with $z=\pm\tfrac{d}{2}$ and arbitrary $x$ and $y$, therefore, in the case when $B_y\neq 0$, the easiest way is to require that the function $\Phi(z)$ has the following properties:
\begin{align}
    \partial_z\Phi(z)\bigl|_{z=\pm\tfrac{d}{2}}\,=\,0\ \ \ \text{and}\ \ \ \Phi(z=\pm\tfrac{d}{2})\,=\,0,
\end{align}
that hold on the scale of the extrapolation length $d= d_{\text{geo}}+\mathcal{O}(1)\xi$.\\

\noindent Dividing Eq.~(\ref{Eq:GL1_separ}) by $\langle\Phi\rangle\neq 0$ and assuming $\langle B_{y/z}\rangle=B_{y/z}$ we get the effective 1st GL equation for a thin 2D film: 
\begin{align}
\label{eq:1stGLeq}
\boxed{0 = -\tfrac{\hbar^2}{4m}(\partial_{xx}+\partial_{yy})\Psi
    -\tfrac{2eB_z}{4m}\hat{L}_z\Psi
    +\tfrac{e^2}{4m}\bigl[B_z^2y^2+(B_z^2 + 4 B_y^2)x^2\bigr]\Psi
    -\bigl[|a(T)|+m\kappa^2 B_y^2\bigr]\Psi
    +\underbrace{\left(b\,\tfrac{\langle\Phi^3\rangle}{\langle\Phi\rangle}\right)}_{b_{\text{eff}}}\,|\Psi|^2\Psi}
\end{align}

We point out that, \textit{in principle}, $B_z$ and $B_y$ alone are sufficient to introduce a certain vortex anisotropy even for $\kappa=0$, as it can be deduced from Eq.~\ref{eq:1stGLeq}. This corresponds to the known effect of vortex axis tilt for superconductors of finite thickness in the presence of Meissner currents, here introduced by the field $B_y$. The shear force exerted by the Meissner currents makes the vortex axis leaning towards the $x$-direction, instead of being parallel to the $z$-axis as in the $B_y=0$ case. The intersection of a tilted cylinder with the $xy$-plane is an ellipse, which is a way to interpret the anisotropy in Eq.~\ref{eq:1stGLeq}. 
Such trivial anisotropy is not relevant for our situation since (i) the thickness in our case is so small (7 nm of Al minus 2 nm of typical oxide) compared to $\xi= 73$~nm,  that the resulting weak screening currents can only produce a very minor tilt; (ii)   even a major tilt could not explain the large squeeze along \textit{both} axes, which is the main results of our work. As explained below, the observed effect requires a finite Lifshitz invariant, namely, a finite $B_y$ and a finite $\kappa$. Experimentally, the necessity of the spin-orbit (i.e. of the Lifshitz invariant) for the observation of the effect is confirmed by our control measurements on the Al/GaAs heterostructure, cf.~gray symbols in Fig.~2a of the main text.

\subsection{Vortex solution $\&$ vortex-core curvatures}

In the ensuing we solve approximately for the vortex the following generic equation of the Gross-Pitaevskii type:
\begin{align}\label{Eq:Vortex_Full}
\boxed{0 = -\mathsf{A}(\partial_{xx}+\partial_{yy})\Psi
    +\tfrac{\mathsf{B}}{i}(x\partial_y-y\partial_x)\Psi
    +\bigl[\mathsf{C}_x\,x^2+\mathsf{C}_y\,y^2\bigr]\Psi
    -\boldsymbol{\alpha}\Psi
    +\boldsymbol{\beta}\,|\Psi|^2\Psi}
\end{align}
where the following coefficients link Eq.~(\ref{eq:1stGLeq}) with Eq.~(\ref{Eq:Vortex_Full}):
\begin{align}\label{Eq:Vortex_Param}
    \mathsf{A}&=\tfrac{\hbar^2}{4m}\,, & 
    \mathsf{C}_x&=\tfrac{e^2B_z^2}{4m}\Bigl(1+4\tfrac{B_y^2}{B_z^2}\Bigr)\,, &
    \boldsymbol{\alpha}&=|a(T)|+m\kappa^2 B_y^2\,, \\
    \mathsf{B}&=-\tfrac{\hbar eB_z}{2m}=\tfrac{\hbar |e|B_z}{2m}\,, &
    \mathsf{C}_y&=\tfrac{e^2B_z^2}{4m}\,, &
    \boldsymbol{\beta}&=b_{\text{eff}}=b\,\tfrac{\langle\Phi^3\rangle}{\langle\Phi\rangle}\,,
\end{align}
Because of a certain similarity of Eq.~(\ref{Eq:Vortex_Full}) with the harmonic oscillator problem (apart from the $\boldsymbol{\beta}$ term) we are interested in a vortex solution in the form:
\begin{equation}\label{Eq:Vortex_Ansatz}
    \Psi(x,y)=K(x+i\,\delta\,y)\,\exp[\tfrac{p}{2}x^2+q xy +\tfrac{r}{2}y^2]\simeq [\text{vortex core region}] \simeq K(x+i\,\delta\,y) + O(x^2,y^2)\,,
\end{equation}
where the approximation on the right side is valid only close to the center of the vortex core. 
The term $(x+i\,\delta\,y)$ defines a vorticity of the solution $\Psi$; for $\delta>0$ ($<0$) it wraps counter-clock-wise
(clock-wise). Moreover, the asymptotic form of our ansatz for $\Psi$ has in the vortex core region the same functional dependence on $x$ and $y$ (including the third powers) as the original Abrikosov solution, see e.g.~Tinkham~\cite{Tinkhambook}.\\

\noindent To obtain an analytical expression for the vortex curvatures, we shall need some further approximations.\\
\textbf{Approximation 1:} We linearize in (\ref{Eq:Vortex_Full}) the non-linear term proportional to $\boldsymbol{\beta}$:
\begin{equation}
    |\Psi(x,y)|^2]\simeq [\text{vortex core region}]\simeq K^2 x^2 + K^2 \delta^2 y^2\propto k_x x^2 + k_y y^2\,,
    \label{eq:psicurv}
\end{equation}
where $k_x\propto K^2$ and $k_y\propto \delta^2 K^2$ are the vortex curvatures along $\hat{x}$ and $\hat{y}$, respectively.
Since we are interested in the vortex core region $\sqrt{x^2+y^2}<\xi\ll \lambda$, we consider only the lowest order terms in $|\Psi|^2$ in Eq.~(\ref{Eq:Vortex_Full}).
Doing so we get linear partial differential equations with the modified quadratic-potential terms:
\begin{align}\label{Eq:Vortex_Approx}
\boxed{0 = -\mathsf{A}(\partial_{xx}+\partial_{yy})\Psi
    +\tfrac{\mathsf{B}}{i}(x\partial_y-y\partial_x)\Psi
    +\bigl[\underbrace{(\mathsf{C}_x+\boldsymbol{\beta}K^2)}_{\mathsf{C}_1}\,x^2
    +\underbrace{(\mathsf{C}_y+\boldsymbol{\beta}K^2\delta^2)}_{\mathsf{C}_2}\,y^2\bigr]\Psi
    -\boldsymbol{\alpha}\Psi}
\end{align}
Plugging the ansatz, Eq.~(\ref{Eq:Vortex_Ansatz}), into the above equation, we get a system of non-linear algebraic equations for the unknown parameters $K,\delta,p,q,r$ (the first two, $K$ and $\delta$, enter also $\mathsf{C}_1$ 
and $\mathsf{C}_2$):
\begin{align}
    0&= \mathsf{C}_1 - \mathsf{A}\, p^2 - i\, \mathsf{B}\, q - \mathsf{A}\, q^2 \label{Eq:Vortex_system1}\\
    0&=\mathsf{C}_2 - \mathsf{A}\, r^2 + i\, \mathsf{B}\, q - \mathsf{A}\, q^2 \\  
    0&= \mathsf{B}\, p\, - \mathsf{B}\, r\,   
    + 2\, i\, \mathsf{A}\,p\, q\,   + 2\, i\, \mathsf{A}\, q\, r\,  \\
     0&=3\, \mathsf{A}\, p + 2\, i\, \mathsf{A}\, q\,\delta + \mathsf{A}\, r - \mathsf{B}\, \delta  + \boldsymbol{\alpha} \\
    0&= 3\, \mathsf{A}\, r\, \delta - 2\,i\, \mathsf{A}\, q\, +\, \mathsf{A}\, p\, \delta - \, \mathsf{B}   
      +\, \delta\, \boldsymbol{\alpha}
      \label{Eq:Vortex_system5}
\end{align}
For convenience of analytical calculation, we change the roles
of $\delta$ and $K$ with $\mathsf{C}_1$ and $\mathsf{C}_2$, i.e., we are assuming as unknowns: $p,q,r,\mathsf{C}_1$, $\mathsf{C}_2$, and as  parameters: $\mathsf{A}, \mathsf{B}, \boldsymbol{\alpha}$, $\delta$.\\

\noindent For a given sign of $\mathsf{B}$, we have the freedom to choose the sign of $\delta$---both enter as parameters in the above equations. However, the basic physics of the superconducting vortices tells us that for $B_z>0$ (implying also $\mathsf{B}>0$) the vortex should wrap counter-clock-wise. This means that in our ansatz, Eq.~(\ref{Eq:Vortex_Ansatz}), we should have for a positive (negative) $B_z$ also positive (negative) $\delta$---this physical requirement implies that combinations like $\mathsf{B}\delta$ or $\mathsf{B}/\delta$ are always positive quantities. 
Solving Eqs.~(\ref{Eq:Vortex_system1})-(\ref{Eq:Vortex_system5}), we obtain two sets of 
solutions---$\mathsf{C}_1^{+}$ and $\mathsf{C}_2^{+}$---and---$\mathsf{C}_1^{-}$ and $\mathsf{C}_2^{-}$---both can be compactly written as follows:
\begin{align}
    \mathsf{C}_1^\pm 
    &= 
    \tfrac{\boldsymbol{\alpha}^2}{128 \mathsf{A}}
    \Bigl[\tfrac{\mathsf{B}^2}{\boldsymbol{\alpha} ^2} \left(\tfrac{9}{\delta ^2}-22 +81 \delta ^2\right)+ 20 
    \pm (6+\tfrac{3\mathsf{B}}{\boldsymbol{\alpha}  \delta }+\tfrac{27 \mathsf{B}\delta}{\boldsymbol{\alpha}  }) \sqrt{\tfrac{\mathsf{B}^2}{\boldsymbol{\alpha} ^2} \left(\tfrac{9}{\delta ^2}-14 +9 \delta ^2\right)+4 \tfrac{\mathsf{B}}{\boldsymbol{\alpha} } \left(\tfrac{1}{\delta }+\delta \right)+4 } + 4 \left(\tfrac{5}{\delta }+9 \delta \right)\tfrac{\mathsf{B}}{\boldsymbol{\alpha} }\Bigr]\nonumber\\
    &\equiv\textcolor{black}{\mathsf{C}_x + \boldsymbol{\beta}K^2} \label{Eq:Vortex-C1}\\
    &{}\nonumber\\
    \mathsf{C}_2^\pm 
    &= 
    \tfrac{\boldsymbol{\alpha}^2}{128 \mathsf{A}}
    \bigl[
    \tfrac{\mathsf{B}^2}{\boldsymbol{\alpha} ^2} \left(\tfrac{81}{\delta ^2}-22 +9 \delta ^2\right)+ 20 
    \pm ( 6 + \tfrac{27\mathsf{B}}{\boldsymbol{\alpha}  \delta } + \tfrac{3 \mathsf{B}\delta}{\boldsymbol{\alpha}  }) \sqrt{\tfrac{\mathsf{B}^2}{\boldsymbol{\alpha} ^2} \left(\tfrac{9}{\delta ^2}-14 +9 \delta ^2\right)+4 \tfrac{\mathsf{B}}{\boldsymbol{\alpha} } \left(\tfrac{1}{\delta }+\delta \right)+4 }
    + 4 \left(\tfrac{9}{\delta }+5 \delta \right)\tfrac{\mathsf{B}}{\boldsymbol{\alpha} }\Bigr]\nonumber\\
    &\equiv\textcolor{black}{\mathsf{C}_y + \boldsymbol{\beta}K^2\delta^2} \label{Eq:Vortex-C2}
    \end{align}
the formulas for $p,q,r$ are not given since we do not need them explicitly---we are looking just for the functional form of $\Psi$ in the first order in $x$ and $y$. We now need to solve Eqs.~(\ref{Eq:Vortex-C1}) and
(\ref{Eq:Vortex-C2}) for $K$ and $\delta$, which is pretty difficult owing to the complicated expressions in square brackets in the above equations. 
Moreover, it is necessary to choose which set of solutions---either ``$+$'': $\mathsf{C}_1^+,\mathsf{C}_2^+$, or 
``$-$'': $\mathsf{C}_1^-,\mathsf{C}_2^-$---is physically more appropriate. This shall be done at the end, when fitting the experimental data.\\

\noindent \textbf{Approximation 2:} To proceed further we expand the lengthy expressions for $\mathsf{C}_1^\pm$, and $\mathsf{C}_2^\pm$ entering Eqs.~(\ref{Eq:Vortex-C1}) 
and (\ref{Eq:Vortex-C2}) assuming:
\begin{equation}
    2\pi\frac{|B_z|\xi^2}{\Phi_0}\approx\Bigl|\frac{\mathsf{B}}{\boldsymbol{\alpha}}\Bigr|\ll 1\,,\ \ \ \ \ \ 
    \frac{\mathsf{B}}{\boldsymbol{\alpha}\delta}\ll 1\,,\ \ \ \ \ \ 
    \text{and}\ \ \ 
    \frac{\mathsf{B}\delta}{\boldsymbol{\alpha}}\ll 1\,.
\end{equation}
By keeping therein terms only up to linear order in $\tfrac{\mathsf{B}}{\boldsymbol{\alpha}}$, $\tfrac{\mathsf{B}}{\boldsymbol{\alpha}\delta}$ and $\tfrac{\mathsf{B}\delta}{\boldsymbol{\alpha}}$, 
a straightforward calculation gives:
\begin{align}
    \mathsf{C}_1^\pm 
    &=\frac{\boldsymbol{\alpha}^2}{4^{1+(0.5-(\pm)0.5)} \mathsf{A}}\Bigl[1+\Bigl(\frac{1}{\delta}\pm 3 \delta \Bigr)\frac{\mathsf{B}}{\mathsf{\boldsymbol{\alpha}}}\Bigr]\,,\label{Eq:VortexApprox-C1}\\
    \mathsf{C}_2^\pm
    &=\frac{\boldsymbol{\alpha}^2}{4^{1+(0.5-(\pm)0.5)} \mathsf{A}}\Bigl[1+\Bigl(\delta\pm \frac{3}{\delta} \Bigr)\frac{\mathsf{B}}{\mathsf{\boldsymbol{\alpha}}}\Bigr]\,.\label{Eq:VortexApprox-C2}
\end{align}
Recalling that $\mathsf{C}^{\pm}_1=\mathsf{C}_x + \boldsymbol{\beta}K^2$ and $\mathsf{C}^{\pm}_2=\mathsf{C}_y + \boldsymbol{\beta}K^2\delta^2$, see Eq.~(\ref{Eq:Vortex_Approx}), the problem reduces for each of the two cases---labeled by ``$+$'' and ``$-$'' sign---to a solution of the algebraic system of two nonlinear equations for two unknowns $K$ and $\delta$: 
\begin{align}
    \mathsf{C}_1^\pm 
    &=\frac{\boldsymbol{\alpha}^2}{4^{1+(0.5-(\pm)0.5)} \mathsf{A}}\Bigl[1+\Bigl(\frac{1}{\delta}\pm 3 \delta \Bigr)\frac{\mathsf{B}}{\mathsf{\boldsymbol{\alpha}}}\Bigr]\, = \mathsf{C}_x + \boldsymbol{\beta}K^2\,,\label{Eq:prefinaleq1}\\
    \mathsf{C}_2^\pm
    &=\frac{\boldsymbol{\alpha}^2}{4^{1+(0.5-(\pm)0.5)} \mathsf{A}}\Bigl[1+\Bigl(\delta\pm \frac{3}{\delta} \Bigr)\frac{\mathsf{B}}{\mathsf{\boldsymbol{\alpha}}}\Bigr]\, = \mathsf{C}_y + \boldsymbol{\beta}K^2\delta^2\,. \label{Eq:prefinaleq2}
\end{align}

\noindent Having found $K$ and $\delta$ the problem is solved, since they provide---apart from a less relevant global scale factor---the curvatures of the vortex core, i.e., $k_x \propto K^2$, and $k_y \propto K^2\delta^2$, see Eq.~(\ref{eq:psicurv}).

\noindent To effectively solve the above systems of equations, it is useful to make explicit the following physical quantities (expressed in terms of the GL parameters $|a(T)|$ and $\boldsymbol{\beta}$)
\begin{align}
    &\text{Bulk Cooper-pair density (per unit volume):} & f_0^2 &=\frac{|a(T)|}{b}=\frac{\langle\Phi^3\rangle}{\langle\Phi\rangle}\frac{|a(T)|}{\boldsymbol{\beta}}\,,\\
    &\text{Bulk condensation energy density (per unit volume):} & e_c^* &=\frac{|a(T)|^2}{b}=|a(T)|f_0^2=b\,f_0^4\,,\\
    &\text{Effective thermodynamic critical field $B_c^*$:} & B_c^* &=\sqrt{2\mu_0 e^*_c}\,,\\
    &\text{GL coherence length:} & \xi &=\frac{\hbar}{\sqrt{4m|a(T)|}}\,,\\
    &\text{Penetration length:} & \lambda &=\sqrt{\frac{m}{2\mu_0 e^2 f_0^2}}\,,\\
    &\text{Lifshitz-Edelstein length:} & \ell_\kappa &=\frac{1}{2\kappa\mu_0 |e| f_0^2}\,,\\
    &\text{Superconducting flux quantum:} & \Phi_0 &=\frac{h}{2|e|}=2\pi\frac{\hbar}{2|e|}\,,
\end{align}
We also use the derived quantities: $m\kappa^2=\tfrac{1}{2\mu_0 f_0^2}\,\tfrac{\lambda^2}{\ell_\kappa^2}$ and $\tfrac{m\kappa}{|e|}=\tfrac{\lambda^2}{\ell_\kappa}$ $\Rightarrow$
\begin{align}
    \frac{\boldsymbol{\alpha}^2}{\mathsf{A}\boldsymbol{\beta}}
    &=
    \frac{\bigl(|a(T)|+m\kappa^2 B_y^2\bigr)^2}{\tfrac{\hbar^2}{4m}\,\boldsymbol{\beta}}
    =
    \frac{\textcolor{black}{|a(T)|}\,\textcolor{black}{|a(T)|}}{\,\textcolor{black}{\tfrac{\hbar^2}{4m}}\,\textcolor{black}{\boldsymbol{\beta}}}
    \Bigl(1+\frac{m\kappa^2}{|a(T)|} B_y^2\Bigr)^2
    =
    \textcolor{black}{\frac{\langle\Phi\rangle}{\langle\Phi^3\rangle}}
    \frac{\textcolor{black}{f_0^2}}{\textcolor{black}{\xi^2}}\,
    \left(
    1+\frac{\lambda^2}{\ell_\kappa^2}\,\frac{1}{e_c^*}\,\frac{B_y^2}{2\mu_0}\right)^2\,,\\
    &\nonumber\\
    \frac{\mathsf{C}_y}{\boldsymbol{\beta}}
    &=
    \frac{e^2 B_z^2}{4m\,\boldsymbol{\beta}}
    =
    \textcolor{black}{\frac{\langle\Phi\rangle}{\langle\Phi^3\rangle}}\frac{1}{4}\,\frac{\textcolor{black}{f_0^2}}{\lambda^2}\,\frac{1}{e_c^*}\,\frac{B_z^2}{2\mu_0}\,,\\
    &\nonumber\\
    \frac{\mathsf{C}_x}{\boldsymbol{\beta}}
    &=
    \textcolor{black}{\frac{\langle\Phi\rangle}{\langle\Phi^3\rangle}}\frac{1}{4}\,\frac{\textcolor{black}{f_0^2}}{\lambda^2}\,\frac{1}{e_c^*}\,\frac{B_z^2}{2\mu_0}\Bigl(1+4\frac{B_y^2}{B_z^2}\Bigr)\,,\\
    &\nonumber\\
    \frac{\boldsymbol{\alpha}}{\boldsymbol{\beta}}\,\frac{\mathsf{B}}{\mathsf{A}}
    &=\textcolor{black}{\frac{\langle\Phi\rangle}{\langle\Phi^3\rangle}}
    \textcolor{black}{f_0^2}\,
    \left(
    1+\frac{\lambda^2}{\ell_\kappa^2}\,\frac{1}{e_c^*}\,\frac{B_y^2}{2\mu_0}\right)\,\frac{2\pi\,B_z}{\Phi_0}\,,
\end{align}

\noindent We now define a dimensionless Cooper pair wave function: $\psi(x,y)=\Psi(x,y)/(\sqrt{\frac{\langle\Phi\rangle}{\langle\Phi^3\rangle}}f_0)$ and expand $|\psi(x,y)|^2$ around the vortex core center
\begin{align}
    |\psi(x,y)|^2 \simeq \tfrac{K^2}{\frac{\langle\Phi\rangle}{\langle\Phi^3\rangle} f_0^2}\,x^2+\tfrac{K^2\delta^2}{\frac{\langle\Phi\rangle}{\langle\Phi^3\rangle} f_0^2}\,y^2
    \equiv
    \textcolor{black}{\frac{1}{2}}k_x x^2 + \textcolor{black}{\frac{1}{2}}k_y y^2.
    \label{eq:kxky}
\end{align}

We have now all the ingredients to compute the curvatures of $|\psi|^2$ near the vortex core. Equation~(\ref{eq:kxky}) links $k_x$ and $k_y$ with $K$ and $\delta$, which in turns are the solutions of the algebraic equations~(\ref{Eq:prefinaleq1})~and~(\ref{Eq:prefinaleq2}). The result below is made explicit for the ``$+$'' case:
\begin{align}
    \frac{1}{\xi^2}\,
    \left(
    1+\frac{\lambda^2}{\ell_\kappa^2}\,\frac{1}{e_c^*}\,\frac{B_y^2}{2\mu_0}\right)^2
    +
    \left(\sqrt{\frac{k_x}{k_y}} + 3 \sqrt{\frac{k_y}{k_x}} \right)\left(
    1+\frac{\lambda^2}{\ell_\kappa^2}\,\frac{1}{e_c^*}\,\frac{B_y^2}{2\mu_0}\right)\,\frac{2\pi\,B_z}{\Phi_0}
    &\equiv
   \frac{1}{\lambda^2}\,\frac{1}{e_c^*}\,\frac{B_z^2}{2\mu_0}\Bigl(1+4\frac{B_y^2}{B_z^2}\Bigr)
    + \textcolor{black}{2} k_x\,,
    \label{Eq:VortexApproxV4-C1}\\
    \frac{1}{\xi^2}\,
    \left(
    1+\frac{\lambda^2}{\ell_\kappa^2}\,\frac{1}{e_c^*}\,\frac{B_y^2}{2\mu_0}\right)^2
    +
    \left(\sqrt{\frac{k_y}{k_x}} + 3\sqrt{\frac{k_x}{k_y}}\right)\left(
    1+\frac{\lambda^2}{\ell_\kappa^2}\,\frac{1}{e_c^*}\,\frac{B_y^2}{2\mu_0}\right)\,\frac{2\pi\,B_z}{\Phi_0}
    &\equiv
    \frac{1}{\lambda^2}\,\frac{1}{e_c^*}\,\frac{B_z^2}{2\mu_0}
    + \textcolor{black}{2} k_y\,,\label{Eq:VortexApproxV4-C2}
\end{align}
since this set of equations turns out to provide a reasonable fit of the experimental data for small in-plane magnetic fields. Due to the significant non-linearity of the problem, it is not, however, excluded that at a certain elevated value of the in-plane field the system can come into a transition point, from which the ``$-$'' case solutions would start to be realized by nature.\\

\noindent \textit{Impact of the crystal structure symmetry.} It is important to stress that the above derivation 
assumes $C_{4v}$ crystallographic symmetry, which is higher than the actual symmetry of InAs-based 2DEG that is possessing the 
crystal symmetry $C_{2v}$. In principle, for a $C_{2v}$-symmetric crystal one might introduce two \textit{different} Lifshitz lengths $\ell_{\kappa1}$ and $\ell_{\kappa2}$ for two different main crystallographic axes---or the so called anisotropic Lifshitz invariant: $\tfrac{1}{2}\kappa_1 B_y (\Psi^*\mathbf{D}_x\Psi+c.c.)+\tfrac{1}{2}\kappa_2 B_x (\Psi^*\mathbf{D}_y\Psi+c.c.)$. 
It turns out, however, that the best fit of our data does not require anisotropic Lifshitz invariant in order to produce the striking curvature enhancement as observed in the experiment. In other words, an isotropic spin-orbit interaction stemming from an isotropic Fermi surface is sufficient to produce an anisotropic vortex squeezing. The vortex anisotropy is exclusively due to the interplay of spin-orbit-coupling with the in-plane magnetic field.



\subsection{Fitting experimental data}
Equations~(\ref{Eq:VortexApproxV4-C1}) and~(\ref{Eq:VortexApproxV4-C2}) are the final output of our model: they implicitly link the curvature along the $x$ and $y$-axis ($k_x$ and $k_y$, respectively) to the Lifshitz length 
$\ell_\kappa$ and to the effective thermodynamic critical field $B_c^*$. 
We have experimentally determined the following constants:
\begin{align}\label{Eq:exp_numbers}
    \xi&=73~\text{nm}\,, & 
    \lambda&=227~\text{nm}\,, &
    B_z&=10~\text{mT}\,, & B_{c2}=61~\text{mT}. 
\end{align}
In particular, $B_{c2}$ is determined as the $B_z$ value (at base temperature and zero bias) for which a resistance emerges such that the RLC circuit resonance is damped. This resistance is, for the present experimental conditions, of the order of 65~$\Omega$~\footnote{At low field, temperature and bias, when the sample inductance is negligible compared to the external inductance of the RLC circuit, the resistance which is sufficient to damp the resonance is just about 1~$\Omega$. However, when the field is increased and approaches the critical value, the sample inductance diverges. Since the $Q$ factor scales as $(\sqrt{L/C})/R$, the increase in sample inductance compensate the effect of the damping resistance. Therefore, near the critical field the resistance at which the resonator becomes overdamped is 65~$\Omega$, which is much larger than 1~$\Omega$, but still much less than the normal resistance.}, which is much less that of half the normal resistance $0.5R_n=0.5\cdot 3074 R_{n,\square}=14.17$~k$\Omega$, where $R_{n,\square}=R_n/3074=9.22$~$\Omega$ is the normal sheet resistance. 
Therefore the $B_{c2}$ value above is slightly less than the conventional value, which is such that  $R(B_z=B_{c2})=0.5R_n$. The value of $\xi$ immediately follows from that of $B_{c2}$ since $B_{c2}=\Phi_0/(2\pi \xi^2)$. The value of $\lambda$ is determined from the sheet kinetic inductance ($L_s/N_{\square}$) at zero field and base temperature, which equals $13$~pH. The sheet kinetic inductance provides directly $\lambda$ via $L_{s,\square}=\mu_0\lambda^2/d$.

We can numerically solve Eqs.~(\ref{Eq:VortexApproxV4-C1}) and~(\ref{Eq:VortexApproxV4-C2}) for $k_x$ and $k_y$ as functions of the in-plane field $B_y$, using the experimentally given values of $\xi$, $\lambda$, and $B_z$, see  Eq.~(\ref{Eq:exp_numbers}), as fixed parameters. The Lifshitz length $\ell_\kappa$ and the effective condensation energy $e_c^{\ast}$---or equivalently $B_c^{\ast}$, since $e_c^{\ast}=(B_c^*)^2/(2\mu_0)$---are taken as the fitting parameters. In order to fit the experimental data displayed in Fig.~2\textbf{a} in the main text, we must relate the vortex curvatures with vortex inductances using  
\begin{align}
 \label{eq:conver}
   L_{v,\perp}=\frac{L_0}{\textcolor{black}{2}\xi^2 k_y}, &\, & L_{v,\parallel}=\frac{L_0}{\textcolor{black}{2}\xi^2 k_x},
\end{align}
where $L_{v,\perp}$ ($L_{v,\parallel}$) is the vortex inductance measured for $\mathbf{B}_{ip}\perp \mathbf{I}$ ($\mathbf{B}_{ip}\parallel \mathbf{I}$). Apart of $\ell_\kappa$ and $B_c^*$ we also fit the conversion parameter $L_0$ 
that serves as a global scaling factor, its theoretical value depends on the microscopic details of the 
pinning strength. 

If we restrict the fitting range of the in-plane field to [-0.1\,T, 0.1\,T] we obtain the solid line curves in Fig.~2\textbf{a} in the main text. The corresponding fitting parameters are: 
\begin{equation}
L_0=2.02\,\text{\textmu H},\ \ \ \ \  
\ell_\kappa=594\,\text{nm},\ \ \ \ \ 
\text{and}\ \ \ \ \ 
B_c^{\ast}=96.1\,\text{mT}.    
\end{equation}
The value of $B_c^*$ is substantially higher than our independent estimate of the thermodynamic critical field $B_c=13$\,mT. This is due to the fact that we have assumed local microscopic in-plane field to be identical to the applied magnetic field. For a few-nm-thick Al film this is in good approximation true (since the thickness is much smaller than both $\lambda$ and $\xi$), however, it is difficult to model the effect of the thicker 2DEG, which also plays a role. 
The discrepancy between local microscopic field and the applied field $B_y$ might be the reason for the difference between the estimated $B_c=13$\,mT and the value of $B_c^{\ast}=96.1$\,mT obtained from the fit. 

We notice that the model nicely captures the prompt decrease of the vortex inductance with the in-plane magnetic field, and the magnitude of the vortex inductance anisotropy. The fit quantitatively reproduces the data up to fields of the order of 100~mT. Above that field range, it systematically overestimates the vortex squeezing effects. A possible explanation for this overestimate is the suppression of the order parameter in the proximitized 2DEG, which recent experiments showed to take place precisely in this  magnetic field range~\cite{Phan2021}.  At the moment this is just an hypothesis, further study is needed to elucidate the relative contribution of 2DEG and Al to the superfluid density, as well as the role of the unconventional pairing to the vortex inductance, as discussed in the main text.

\section{Inductance per vortex in the linear regime of $L_v(B_z)$}
Figure~1\textbf{c} of the main text shows the dependence of the vortex inductance $L_v$ on the out-of-plane field $B_z$. In particular, the inset makes it evident that for moderate fields (up to 20~mT) the behavior is linear, i.e., each vortex added into the system contributes with the same additional inductance. This is what one would expect if the interaction between vortices is negligible. This is the case if the separation between votices is much larger then $\xi$ or, equivalently, if $B_z\ll B_{c2}$. Eventually, when $B_z$ becomes a significant fraction of $B_{c2}$ (in our case for $B_z=20$~mT$\approx B_{c2}/3$) then neighboring cores start overlapping and each new vortex also contributes to the reduction of the superfluid density in its vicinity, producing a superlinear increase in the vortex inductance.

It is interesting to verify whether the measured slope  $L_v/B_z = 118$\,nH/mT  is compatible with Eq.~1 of the main text, with a reasonable assumption for $k$. From the theory we know that  $k \approx 0.25 d B_c^2/\mu_0$~\cite{Golosovsky1996,BlatterRMP}, where  $B_c=13.9$\,mT is the thermodynamic critical field and $d\simeq 4.5\,$nm~\cite{Wang2020} the effective Al thickness, i.e. the nominal one minus 2.5~nm of oxide. With this value of $k$, using Eq.~1 we obtain $L_v/B_z=N_{\square}{\Phi_0}/{k}=36$\,nH/mT, which is of the same order of magnitude as the measured value. The factor three discrepancy between the expected and the measured value of $L_v/B_z$ is acceptable, in particular when considering the relatively large theoretical uncertainty for the numerical prefactor 0.25~\cite{Golosovsky1996}. Also, our synthetic Rashba superconductor has a complex structure along the $z$ direction (a metallic film, an insulating barrier and a proximitized 2DEG), which is clearly not considered in the simple models.


\section{Isotropy of the kinetic inductance in the absence of vortices}
\begin{figure}[tb]
	\includegraphics[width=\textwidth]{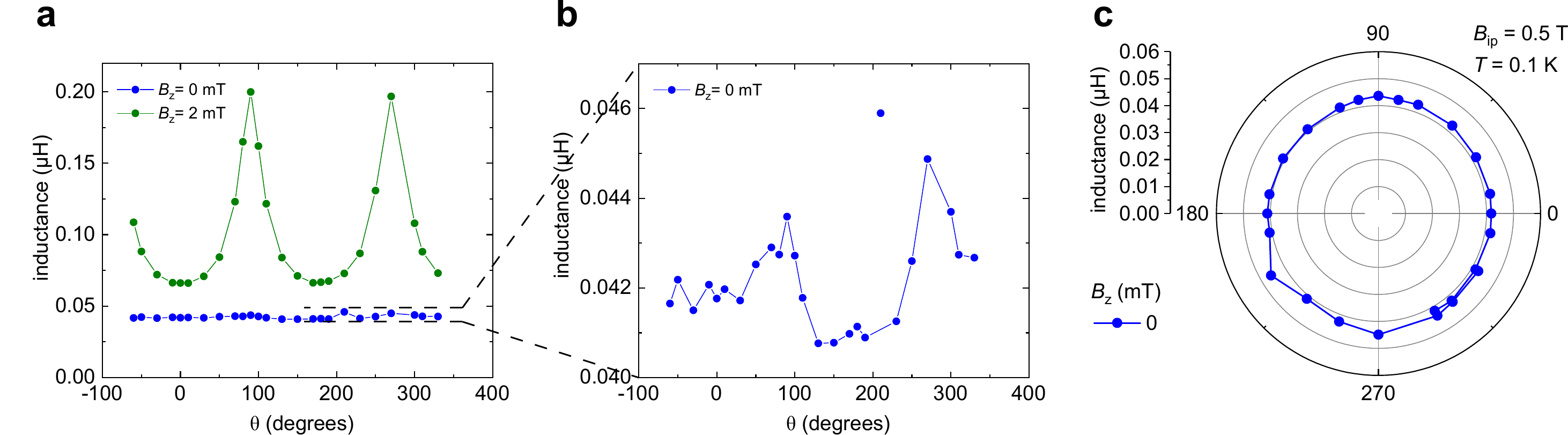}
	\caption{\textbf{a,} Inductance measured for $B_z=0$~mT (blue) and $B_z=2$~mT (green) as a function of the angle $\theta$ between in-plane field $\vec{B}_{ip}$ and current $\vec{I}$, for $|\vec{B}_{ip}|=$0.5~T and $T=0.1$~K. 
	 The graph corresponds to that in Fig.~2\textbf{b} of the main text, displayed in a cartesian plot. \textbf{b,}  Zoom in order to highlight the data for $B_z=0$.The outlier is most probably caused by a flux jump in the compensation coil. \textbf{c,} Same data as in panel \textbf{b}, but displayed in a polar plot.
	}
	\label{fig:isotropicBzero}
\end{figure}

Figure~\ref{fig:isotropicBzero} focuses on the inductance measurements at $B_{ip}=0.5$~T for $B_z=0$~mT (blue) and $B_z=2$~mT (green). The former case corresponds, in good approximation, to the absence of vortices, while the latter case to a small but finite vortex density. In this section we show how different is the symmetry of the inductance measured in the two cases.

Panel \textbf{c} shows a magnified view of the inductance curve for the $B_z=0$ case in Fig.~2\textbf{b} of the main text. This nearly isotropic graph must be compared to the highly anisotropic ones at finite $B_z$ in Fig.~2\textbf{b}. For ease of comparison we displayed both the graph for $B_z=0$ (blue) and for $B_z=2$~mT (green) in a cartesian plot, see Fig.~~\ref{fig:isotropicBzero}\textbf{a}. The residual anisotropy, barely discernible in the zoom-in plot displayed in panel~\textbf{b}, might be due to residual vortices. In fact, the orthogonal coils used for compensating $B_z$ (the compensation procedure is outlined in the last section) produce a field which is not \textit{perfectly} homogeneous. As a consequence, if the sample is large, it is possible to locally have uncompensated vortices, even when the average $B_z$ is globally set to zero. Since vortices provide a much larger inductance contribution than the bare kinetic inductance, their effect can be important. Therefore, Fig.~\ref{fig:isotropicBzero} sets an \textit{upper limit} for the anisotropy of the bare kinetic inductance, which is evidently very low. 

From Fig.~\ref{fig:isotropicBzero} we deduce that, at least in good approximation, the superfluid density is  isotropic even when subjected to in-plane fields.
In the absence of Lifshitz invariant terms in the free energy, one would then expect a similar isotropy also for the vortex core structure (and correspondingly for the measured vortex inductance). The observed strong anisotropy of $L_v$ for finite $B_z$ and $B_{ip}$ is a  strong signature of the Lifshitz invariant. This is further corroborated by control measurements in samples with largely reduced SOI (see next section), where both pinning enhancement and inductance  anisotropy are \textit{not} observed.

\clearpage
\section{Isotropy of kinetic inductance and vortex inductance in $\mathbf{Al/GaAs}$ samples}

\begin{figure}[htb]
	\includegraphics[width=0.7\textwidth]{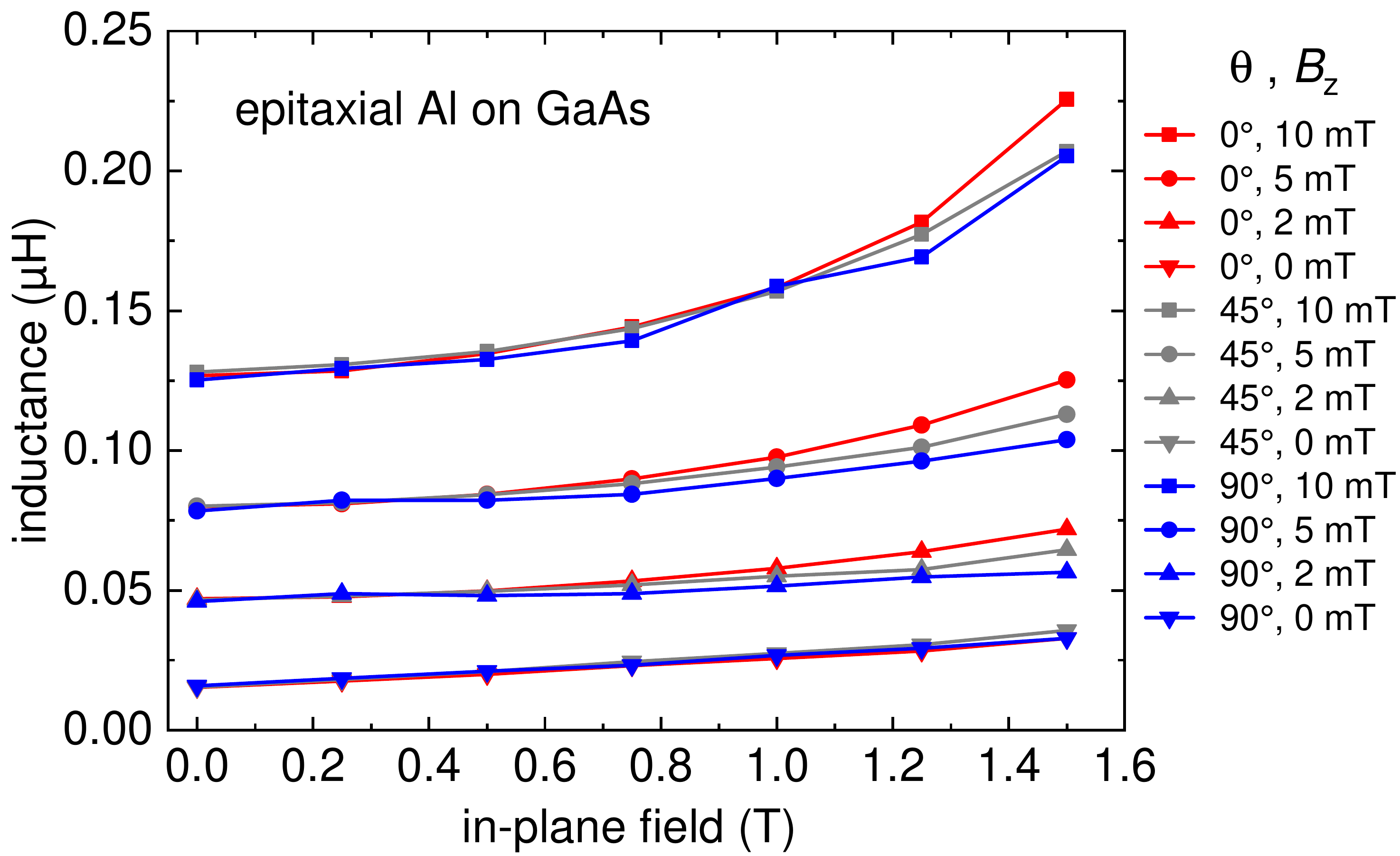}
	\caption{Inductance measured on a meander device patterned on a Al/GaAs sample, where spin-orbit effects are largely reduced. Measurements are performed for different angles $\theta$ between in-plane field $\vec{B}_{ip}$ and current $\vec{I}$ [$\theta=0^{\circ}$ (red), $\theta=45^{\circ}$ (grey), $\theta=90^{\circ}$ (blue)], for different values of $B_z$ [0~mT ($\bigtriangledown$), 2~mT ($\bigtriangleup$), 5~mT ($\square$), 10~mT ($\bigcirc$)] and of $B_{ip}$ [abscissas]. We notice that the inductance always monotonically increases. The anisotropy in the inductance is negligible: it can only be discerned for finite $B_z$ (vortex inductance) and large $B_{ip}$ (larger than 1~T). 
	}
	\label{fig:isogaas}
\end{figure}

Figure~\ref{fig:isogaas} shows the results of inductance measurements performed on our control sample. As mentioned in the main text, this sample is a meander structure similar to the main device discussed in this work, see Fig.~1\textbf{b} of the main text.
This meander is patterned starting from a heterostructure consisting of Al grown on top of a GaAs substrate. While the Al film is similar to that grown on InAs, the absence of a quantum well, together with the reduced atomic weight of Ga compared to In, guarantees that SOI is greatly reduced.

The graph in Fig.~\ref{fig:isogaas} shows the inductance measured on such control device for selected values of $B_{ip}$, $B_z$ and $\theta$. For $B_{ip}<1$~T the inductance is nicely isotropic. Also, as anticipated in the main text (Fig.~2\textbf{a}, grey symbols) the vortex inductance increases with $B_{ip}$.
Only at very large $B_{ip}$ \textit{and} at finite $B_z$, a slight anisotropy emerges. Such anisotropy is comparable with the data point scatter and it is much smaller than the vortex anisotropy observed in the epitaxial Al/InAs sample discussed in the main text. 
Data in Fig.~\ref{fig:isogaas} unambiguously demonstrate the crucial role of the SOI in determining the observed anisotropic vortex squeezing.

\section{Weak anisotropy for $B_{c,ip}$}
\begin{figure}[htb]
	\includegraphics[width=0.6\textwidth]{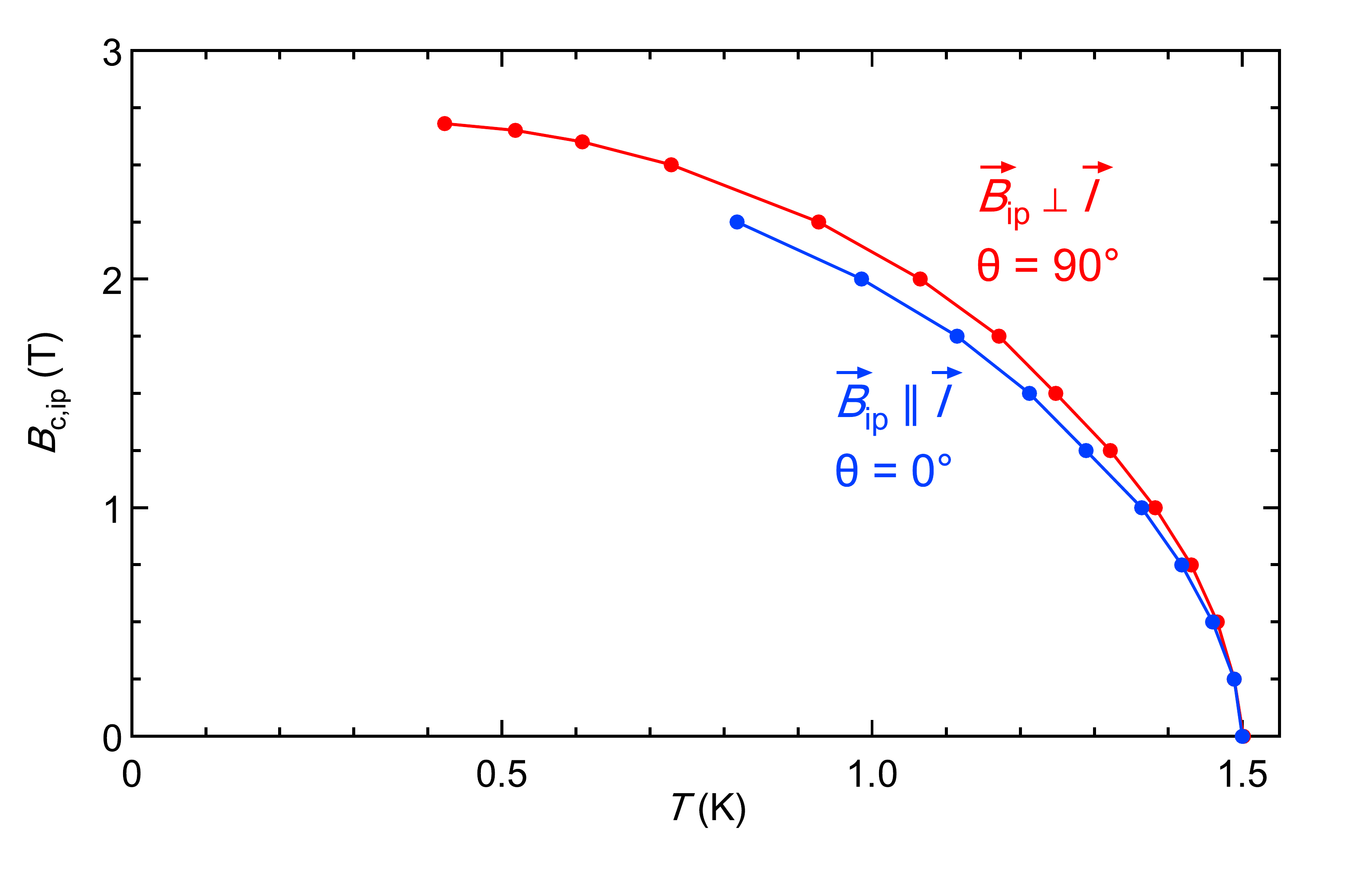}
	\caption{The graph shows the critical in-plane magnetic field $B_{c,ip}$ as a function of the temperature $T$ for $\theta=90^{\circ}$ (red, $\mathbf{B}_{ip}\perp \mathbf{I}$) and  $\theta=0^{\circ}$ (blue, $\mathbf{B}_{ip}\parallel \mathbf{I}$).
	}
	\label{fig:anisoBcpar}
\end{figure}

Figure~\ref{fig:anisoBcpar} shows the temperature dependence of the in-plane critical magnetic  field $B_{c,ip}$. The critical field values correspond to the emergence of a resistance $R(B_{c,ip})=0.5R_n$ where the normal state resistance $R_n=9.2$~$\Omega$. 
The measurement has been repeated for $\theta=0^{\circ}$ (blue curve, $\mathbf{B}_{ip}$ parallel to the current) and for $\theta=90^{\circ}$ (red curve, $\mathbf{B}_{ip}$ perpendicular to the current).

We notice a small anisotropy (of the order of 8\%) in $B_{c,ip}$, which implies the same anisotropy for $\xi \propto B_{c,ip}$. Since the width and thus the curvature of the potential $U(\mathbf{r})$ scale as $\xi$, in principle the anisotropy in $\xi$ should determine an anisotropy in the measured vortex inductance. From a quantitative point of view, however, the observed $\xi$ anisotropy is too small to justify a large difference between $L_v(\theta=0^{\circ})$ and   $L_v(\theta=90^{\circ})$. In fact, the ratio $L_v(\theta=0^{\circ})/L_v(\theta=90^{\circ})$ (i.e. the ratio $k_{\perp}/k_{\parallel}$) is about 4.7 at $B_{ip}=1$~T, see Fig.~2\textbf{c} of the main text.

What we learn from the small anisotropy in the in-plane critical field $B_{c,ip}$ (Fig.~\ref{fig:anisoBcpar}) and in the kinetic inductance $L_s$ (Fig.~\ref{fig:isotropicBzero}) is that the presence of $\mathbf{B}_{ip}$ does impact the isotropy of the condensate (as highlighted by recent studies~\cite{Phan2021}), but this effect is small and thus insufficient to explain the strong anisotropy of the vortex inductance. On the other hand, the vortex anisotropy  naturally emerges as a result of the Lifshitz invariant term in the Ginzburg-Landau free energy.

\section{Vortex inductance for low in-plane fields: zooming in Fig.~2$\mathbf{a}$}
\begin{figure}[hbt]

	\includegraphics[width=0.6\textwidth]{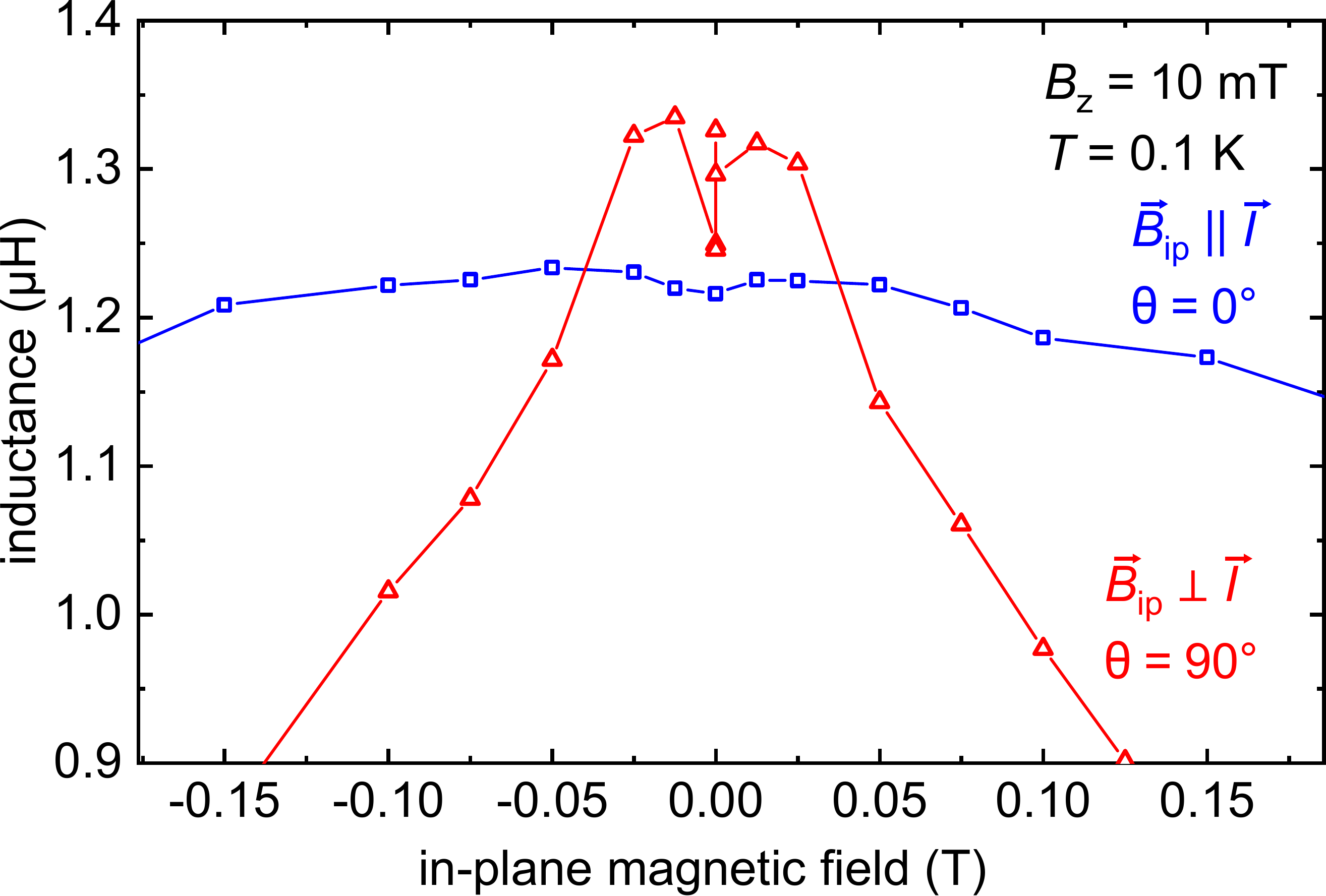}
	\caption{Zoom-in of Fig.~2\textbf{a} of the main text. 
	}
	\label{fig:zoomin2a}
\end{figure}

Figure~\ref{fig:zoomin2a} shows a zoom-in view of the low in-plane field region in Fig.~2\textbf{a} of the main text. The graph makes it evident that what in the full range graph in Fig.~2\textbf{a} appeared as noise near zero field, is indeed a double peak with a cusp-like minimum at zero, where the values for $\theta=90^{\circ}$ and $\theta=0^{\circ}$ become approximately equal. 
This double-peak is visible also for the  $\theta=0^{\circ}$ data (blue in Fig.~\ref{fig:zoomin2a}), although in a less pronounced fashion.

The study of these peaks is beyond the scope of this article.  Investigation on these peaks is ongoing, and it will be discussed elsewhere.  

\section{Further measurements of the depinning current versus in-plane field}

\begin{figure}[tb]
	\includegraphics[width=\textwidth]{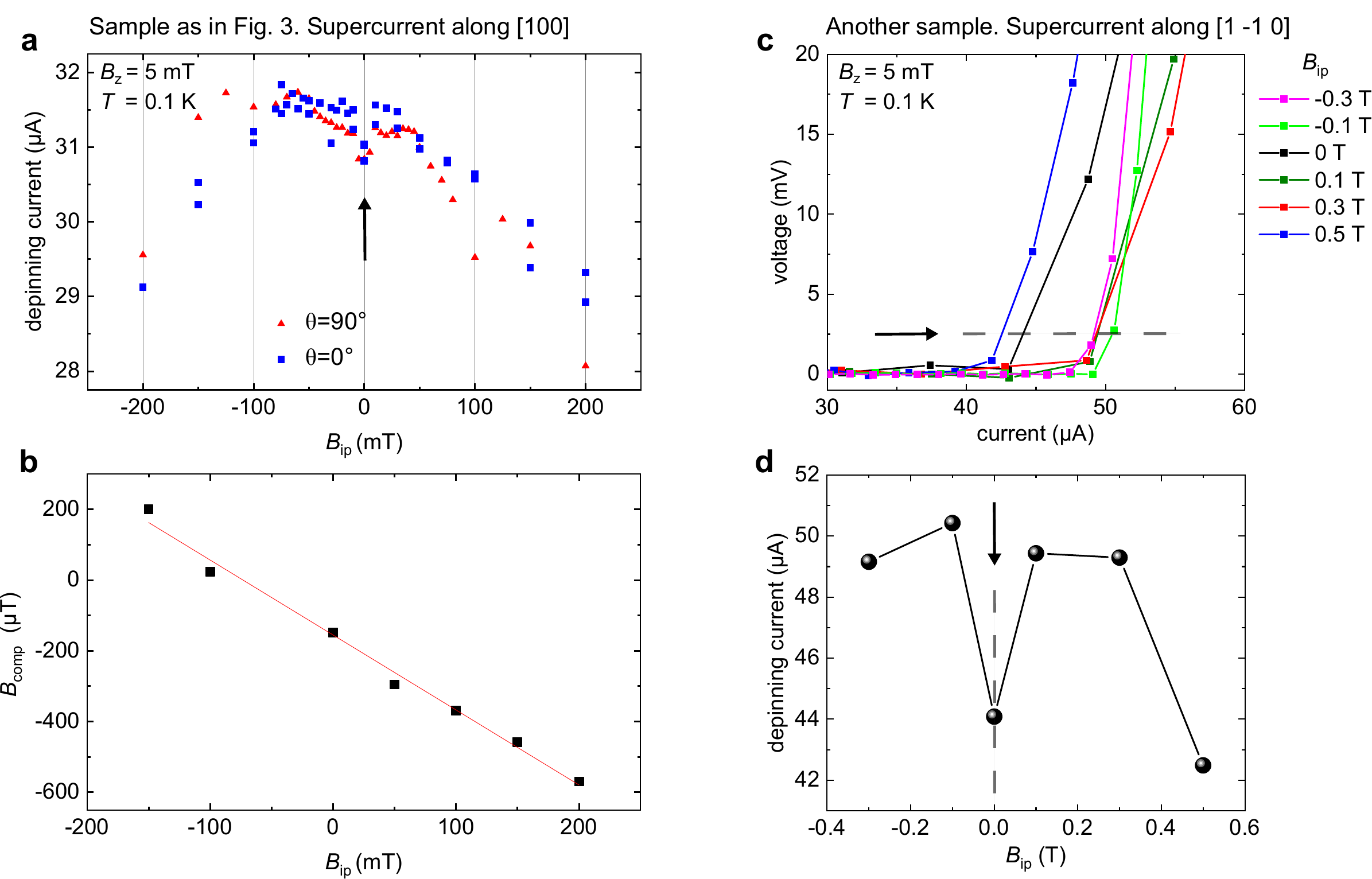}
	\caption{\textbf{a,} Depinning current versus in-plane field $B_{ip}$, measured at $B_z=5$~mT for in-plane field perpendicular (red symbols $\theta=90^{\circ}$) and parallel (blue symbols $\theta=0^{\circ}$) to the current. The measurement is performed on the same sample discussed in Fig.~3 of the main text, where the current is directed along the [100] crystallographic direction. The arrow indicates the anomalous minimum at zero field.  \textbf{b,} Out-of-plane field $B_{comp}(B_{ip})$ that must be applied to obtain an effective zero out-of-plane field $B_z$ in the measurements in panel \textbf{a}. The finite slope originates from the misalignment of $\mathbf{B}_{ip}$, which is not perfectly perpendicular to $B_z$.  \textbf{c,} IV-characteristics measured on a different device, with the same geometry as the one of the previous panels, but oriented along the [1-10] crystallographic direction. The arrow indicates the threshold (2.5 \textmu V) used to determine the depinning current. \textbf{d,} Depinning current  as a function of the in-plane field for the latter sample. A minimum at zero field is again visible, indicated by the arrow.
	}
	\label{fig:combo}
\end{figure}

In this section we shall discuss further measurements of the in-plane field dependence of the depinning current. These measurements were performed with different orientations of the in-plane magnetic field with respect to the current ($\theta=0^{\circ}$ and  $\theta=90^{\circ}$). At the end of the section, we also present measurements on a different device.

Figure~\ref{fig:combo}\textbf{a} shows the same measurements reported in Fig.~3 of the main text (red symbols, $\theta=90^{\circ}$), together with the same measurements performed with the in-plane field oriented parallel to the current (blue symbols, $\theta=0^{\circ}$). We notice that the minimum of the depinning current at zero bias (black arrow) is reproducible. The minimum for $\theta=0^{\circ}$ is less pronounced compared to that for $\theta=90^{\circ}$, which possibly mirrors the fact that the anomalous inductance decrease is less pronounced for $\theta=0^{\circ}$ compared to $\theta=90^{\circ}$, see Fig.~2 of the main text.

We stress that the depinning current measurements here shown were much more difficult compared to inductance measurements, owing to the fact that the devices we used had a very large contact resistance. These required the use of very fast (9~ms) IV sweeps, followed by long cooling times (30~s). More importantly, the large resistance made it impossible to make use of the compensation field routine described in the next section, which we followed for the inductance measurements. In this case we followed a different routine, described in the following:
\begin{enumerate}
    \item The desired in-plane field $B_{ip}$ is set.
    \item The $B_z$ field component is set to the expected zero value.
    \item The sample is heated above $T_c$, then a waiting time of 15 minutes allows the sample to cool down to base temperature.
    \item A series of of IV-traces is taken for different $B_z$ values. The set $B_z=B_{comp}$ value such that critical current is maximal is taken as effective $B_z=0$ value.
    \item The operations above are repeated for several in-plane field values $B_{ip}$ that cover the desired range of values for the final measurements. As a result, one obtains a graph as the one in Fig.~\ref{fig:combo}\textbf{b}, showing the necessary $B_{comp}(B_{ip})$ component necessary to compensate the each  $B_{ip}$.
\end{enumerate}
After $B_{comp}(B_{ip})$ is determined, we start the final measurement of the depinning current versus $B_{ip}$, by applying for each  $B_{ip}$ the corresponding compensation field plus 5~mT, i.e.,  $B_z^{comp}(B_{ip})+5$~mT. Then, the sample is heated above $T_c$, cooled down back to base temperature, then finally 45 IVs are measured as discussed in the main text.
As an alternative, we applied the compensation routine at each $B_{ip}$ point. This is the case of the $\theta=0^{\circ}$ points (blue symbols) in Fig.~\ref{fig:combo}\textbf{a}. However, this makes the entire measurement series much longer, with an increased risk of drift of important measurement parameters.

In Fig.~\ref{fig:combo}\textbf{c,d} we show depinning current measurements versus in-plane field for another device, with exactly the same geometry as the one discussed above. In this case, however, the current (i.e., the axis of the Hall bar) is directed along the [1-10] crystallographic direction. Again, each depinning current value is determined after averaging of many fast IV-characteristics. The orientation of the in-plane field is $\theta=90^{\circ}$, while the out-of-plane $B_z=5$~mT. The panel \textbf{c} shows selected IV characteristics, while panel \textbf{d} shows the deduced depinning current values as a function of $B_y$. Again, we observe a clear minimum at zero-field, consistent with our other observations reported above and with the inductance measurements in the main text. As for the anomalous inductance decrease discussed in the Fig.~2 of the main text, this anomalous increase of the pinning strength is a clear signature of the impact of the Lifshitz invariant on the condensate.

\section{Out-of-plane field compensation procedure}
\begin{figure}[htb]
	\includegraphics[width=0.6\textwidth]{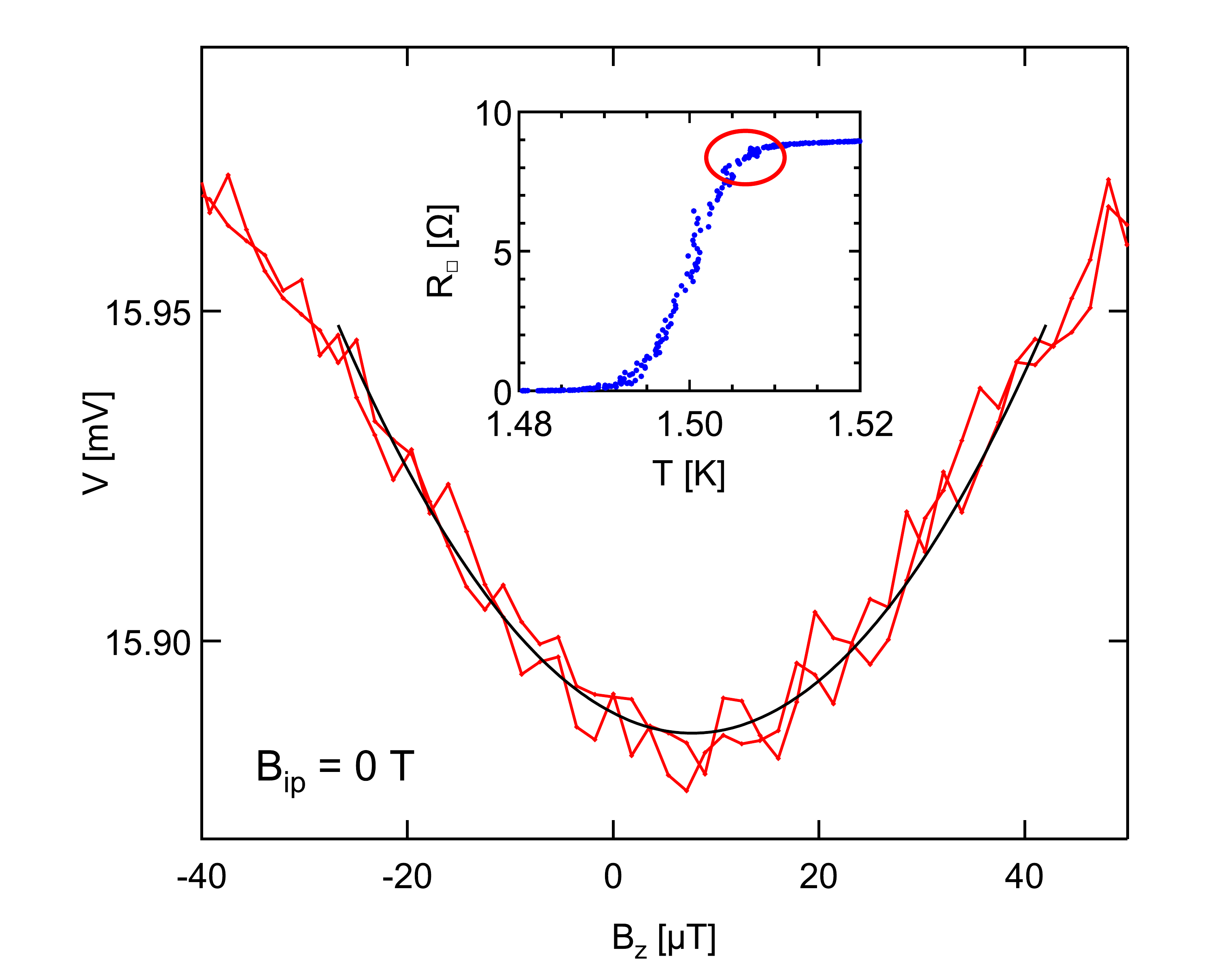}
	\caption{Exemplary magnetoresistance measured at $T>T_c$. Data are taken at a temperature corresponding to the fluctuation regime of the superconductor, where $R(B_z)$ is non-hysteretic. This regime is highlighted in the inset. Black solid line is a parabolic fit to the measured data.
	}
	\label{fig:compensation_example}
\end{figure}

Owing to misalignment of the different parts of the cryostat, the sample surface will hardly be \textit{perfectly}  parallel to the axis of the main superconducting coil, which we use to apply the in-plane field $\mathbf{B}_{ip}$. 
Owing to the large inductive contribution of vortices, the undesired  out-of-plane component of $\mathbf{B}_{ip}$ will have a significant impact on the inductance measurements. In particular, it will mask the measurement of the kinetic inductance of the condensate.   
For measurements that demand zero out-of-plane magnetic field, residual out-of-plane field components must be manually compensated using additional orthogonal coils. In this section we outline the procedure we used to zero the out-of-plane field in inductance measurements. The ideas is to identify a physical quantity which is very sensitive to the absolute  perpendicular fields, as e.g.~the resistance near the superconducting transition near the critical temperature. In Fig.~\ref{fig:compensation_example} a typical compensation measurement is depicted. In this case, the resistance of the sample slightly above the critical temperature (see inset of Fig.~\ref{fig:compensation_example}) is measured. In this regime, the resistance is strongly dependent on the perpendicular field, which is applied via the additional orthogonal coils. Perpendicular field sweeps are always performed back and forth to ensure that the resistive response is non-hysteretic in order to exclude e.g.~flux trapping. 
The compensation field is determined by finding the minimum of a parabolic fit of the magnetoresistance data measured in the fluctuation regime of the superconductor. In that regime surface barriers for vortex entry and exit are absent and hence magnetoresistance curves are non-hysteretic. With this procedure the compensation field can be easily determined with an uncertainty below  $10$~$\mathrm{\mu}$T.

The perpendicular field $B_{comp}$ needed to compensate for the sample misalignment is therefore
\begin{equation}
	B_{comp}(B_{ip}) = B_{ip} \sin(\alpha) ,
	\label{eq:comp_angle}
\end{equation}
where $\alpha$ is the angle between the film plane and the applied in-plane field $\mathbf{B}_{ip}$. The compensation field is clearly proportional to $B_{ip}$. Owing to the imperfect homogeneity of the compensation field, together with the fact that the sample position within the field distribution changes with $\theta$  (since the rotation axis of the piezo is not perfectly centered on the sample), the misalignment angle $\alpha$ also depends on $\theta$. Thre angle $\alpha$ is typically less than $2{^\circ}$. 

\clearpage

\end{document}